\documentclass[preprint,12pt,authoryear]{elsarticle}

\usepackage{amssymb}
\usepackage{amsmath}
\usepackage{hyperref}

\usepackage{color, listings, setspace, courier, relsize}

\definecolor{Code}{rgb}{0,0,0}
\definecolor{Decorators}{rgb}{0.5,0.5,0.5}
\definecolor{Numbers}{rgb}{0.5,0,0}
\definecolor{MatchingBrackets}{rgb}{0.25,0.5,0.5}
\definecolor{Keywords}{rgb}{0,0,1}
\definecolor{self}{rgb}{0,0,0}
\definecolor{Strings}{rgb}{0,0.63,0}
\definecolor{Comments}{rgb}{0,0.63,1}
\definecolor{Backquotes}{rgb}{0,0,0}
\definecolor{Classname}{rgb}{0,0,0}
\definecolor{FunctionName}{rgb}{0,0,0}
\definecolor{Operators}{rgb}{0,0,0}
\definecolor{Background}{rgb}{0.98,0.98,0.98}
\definecolor{Booleans}{rgb}{0.572,0,0.572}
\definecolor{BuiltinFunction}{rgb}{0.572,0,0.572}
\definecolor{BuiltinConstant}{rgb}{0.572,0,0.572}
\definecolor{Asterisk}{rgb}{0.670,0,1}

\lstdefinelanguage{Python}{
    	numbers=left,
    	numberstyle=\footnotesize,
    	numbersep=7pt,
    	xleftmargin=1.26em,
    	framextopmargin=2em,
    	framexbottommargin=2em,
    	showspaces=false,
    	showtabs=false,
    	showstringspaces=false,
    	frame=l,
    	tabsize=4,
    	stepnumber=1,
	basicstyle=\ttfamily\small,
    	backgroundcolor=\color{Background},
    	breaklines=True,
    	postbreak=\mbox{\textcolor{red}{$\hookrightarrow$}\space},
	commentstyle=\color{green}\ttfamily,
    	stringstyle=\ttfamily\color{Strings},
    	morecomment=[s][\color{Strings}]{'}{'}, 
    	stringstyle=\ttfamily\color{Comments},
    	morecomment=[s][\color{Comments}]{\#}{\#}, 			
	stringstyle=\ttfamily\color{Strings},
	morekeywords={import,from,class,def,for,while,if,is,in,elif,else,not,and,or,print,break,continue,return,access,as,except,exec,finally,global,import,lambda,pass,print,raise,try,assert},
    	keywordstyle={\color{Keywords}\bfseries}, 
    	morekeywords={[2]True,False,None},
    	keywordstyle={[2]\color{BuiltinConstant}\slshape},
	otherkeywords={[2]*},
	keywordstyle={[2]\color{Asterisk}},
	emph={self},
	emphstyle={\color{self}\slshape}	
}
   
\lstdefinelanguage{bash}{
    numbers=none,
    numberstyle=\footnotesize,
    numbersep=7pt,
    xleftmargin=1.26em,
    framextopmargin=2em,
    framexbottommargin=2em,
    showspaces=false,
    showtabs=false,
    showstringspaces=false,
    frame=none,
    tabsize=4,
    stepnumber=1,
    basicstyle=\ttfamily\small\setstretch{1},
    backgroundcolor=\color{Background},
    breaklines=True,
    postbreak=\mbox{\textcolor{red}{$\hookrightarrow$}\space},
}    

\newsavebox{\LstBox}

\newcommand{\pythonpackage}{\texttt{lenstronomy}}
\newcommand{\pythonpackageGitHub}{\url{https://github.com/sibirrer/lenstronomy}}
\newcommand{\pythonpackagePyPi}{\url{https://pypi.python.org/pypi/lenstronomy}}
\newcommand{\docLink}{\url{https://lenstronomy.readthedocs.io}}
\newcommand{\tutorialLink}{\url{https://github.com/sibirrer/lenstronomy_extensions}}
\newcommand{\coverallLink}{\url{https://coveralls.io/github/sibirrer/lenstronomy}}
\newcommand{\travisLink}{\url{https://travis-ci.org/sibirrer/lenstronomy}}

\newcommand{\Python}{\texttt{python}}
\newcommand{\pythonPSL}{\url{https://docs.python.org/2/library/index.html}}

\journal{Physics of the Dark Universe}

\begin{document}

\begin{frontmatter}



\title{\pythonpackage: multi-purpose gravitational lens modelling software package}


\author{Simon Birrer\corref{cor1}}
\cortext[cor1]{corresponding author and lead developer}
\ead{sibirrer@astro.ucla.edu}
\address{Department of Physics and Astronomy, University of California, Los Angeles, 475 Portola Plaza, Los Angeles, CA 90095-1547, USA}

\author{Adam Amara\corref{cor2}}
\ead{adam.amara@phys.ethz.ch}
\address{Institute of Particle Physics and Astrophysics, Department of Physics, ETH Zurich, Wolfgang-Pauli-Strasse 27, 8093, Zurich, Switzerland}

\begin{abstract}
We present \pythonpackage, a multi-purpose open-source gravitational lens modeling \Python\ package. \pythonpackage\ is able to reconstruct the lens mass and surface brightness distributions of strong lensing systems using forward modelling. \pythonpackage\ supports a wide range of analytic lens and light models in arbitrary combination. The software is also able to reconstruct complex extended sources (Birrer et. al 2015) as well as being able to model point sources. We designed \pythonpackage\ to be stable, flexible and numerically accurate, with a clear user interface that could be deployed across different platforms. Throughout its development, we have actively used \pythonpackage\ to make several measurements including deriving constraints on dark matter properties in strong lenses, measuring the expansion history of the universe with time-delay cosmography, measuring cosmic shear with Einstein rings and decomposing quasar and host galaxy light. The software is distributed under the MIT license. The documentation, starter guide, example notebooks, source code and installation guidelines can be found at \docLink.
\end{abstract}


\end{frontmatter}


\section{Introduction} \label{sec:introduction}
Strong gravitational lensing, the bending of light by foreground masses to such an extent that multiple images of the same source are formed, is an important phenomenon that can be used to probe the matter distribution and geometry of the universe. The detailed reconstruction of the light paths can be used to test the nature of the unknown components dark matter and dark energy. These dominate the matter-energy content of the Universe today.

In strong lensing studies, significant progress has been made in recent years in both quantifying the small scale matter distribution, with techniques such as gravitational imaging \cite{Vegetti:2010mb, Vegetti:2012au, Hezaveh:2016uj, Birrer:2017a, Vegetti:2018} and flux ratio anomalies \cite{Mao_Schneider:1998, Metcalf_Madau:2001, Dalal:2002, Nierenberg:2014, Xu:2015, Nierenberg:2017}, and measuring the expansion history of the universe, with time-delay cosmography \cite{Refsdal:1964pi, Schechter:1997, Treu_Koopmans:2002, Suyu:2010rc, Suyu:2014aq, Birrer2016_mst, Bonvin:2017}. These successes have in part been made possible due to the development of state-of-the-art lens modelling software and algorithms that can extract the required lensing information from high resolution imaging data. Several of the codes used for doing this have been made publicly available to the community (see \ref{app:software}).

From the current and upcoming surveys, the sample of known strong lenses is rapidly increasing \citep[see e.g.][]{Agnello:2015DES, Nord:2016, Schechter:2017, Lin:2017DES, Jacobs:2017, Ostrovski:2017, Williams:2017, Lemon:2018} and Treu et al. (submitted).
This enlarged sample enables competitive measurements of the Hubble constant \citep[see e.g.][]{Treu:2016, Shajib:2018, Suyu:2018} as well as can strengthen constraints on dark matter properties on sub-galactic scales \citep[see e.g. forecast by][]{Gilman:20117_abc}. To fully exploit the science potential of these new strong lensing data, our modelling tools need to be continually developed and improved. 

In this publication, we present the first public release of \pythonpackage, which is an open source multi-purpose strong lens modelling software package. \pythonpackage\ has been used as a research tool throughout its development. This includes being used for time-delay cosmography \citep{Birrer2016_mst, Birrer:J1206}, lensing substructure analysis \citep{Birrer:2017a}, line-of-sight shear measurements from an Einstein ring \citep{Birrer_2017los} and its forecast to measure cosmic shear with Einstein rings \citep{Birrer2018_cosmic_shear}. Each of these tasks required the modelling of \textit{Hubble Space Telescope} (HST) imaging data at the pixel level.

We developed \pythonpackage\ to be stable, flexible and numerically accurate, with a clear Application programming interface (API) that could be used across different platforms. The \pythonpackage\ software architecture was designed to be able to scale from the current era, where individual lenses are studied in detail, to the case where several hundreds of lenses, from future surveys, will need to be processed.  

Given that an essential part of precision cosmology is the control of systematic errors, it is important for the community to develop multiple independent pipelines. This allows for the cross-checks that are necessary for complex precision measurements. Community standard benchmark efforts also make important contributions. For instance, the time-delay lens modelling challenge \citep{tdlmc:2018} offers a realistic and blind comparison framework in the domain of cosmography. Similar efforts are underway by the substructure lensing community. The public release of \pythonpackage\ enables a transparent and effective comparison with other software used in strong lensing.

\pythonpackage\ includes, but is not limited to, the methods presented in \citep{Birrer2015_basis_set}. This includes a linear source reconstruction method based on Shapelet \citep{Refregier:2003eg} basis sets, a Particle Swarm Optimization \citep{Eberhart:1995qm} for optimizing the non-linear lens model parameters and a MCMC framework for Bayesian parameter inference \citep[\texttt{emcee}][]{emcee}. The software supports a high dynamic range in angular scales, complexity in source and lens models, can handle various image qualities and meets the requirements for diverse science applications. Furthermore, \pythonpackage\ enables a consistent integration of imaging, time-delay and kinematic data to provide model constraints.

There is continued development and support of \pythonpackage\ to expand its scope and scientific application for a growing user community.

This paper is structured as follows: In Section \ref{sec:overview}, we provide an overview of the software architecture and deployment, including installation and dependencies. In Section \ref{sec:core_modules}, we describe the core modules of \pythonpackage\ with some simple examples of how to use them. We provide some modelling examples in Section \ref{sec:model_examples} to demonstrate the capabilities and flexibility of \pythonpackage. In Section \ref{sec:applications}, we give some science application highlights. We summarize in Section \ref{sec:conclusion}.

This publication is accompanied by a public release of the source code with extensive online documentation\footnote{\docLink}. Additionally, we provide \texttt{Jupyter}\footnote{\url{http://jupyter.org}} notebook examples for different science cases. We refer the reader to these online resources for the most updated version of the software package.

\section{Package overview}
\label{sec:overview}

\begin{lrbox}{\LstBox}
\begin{lstlisting}[language=bash]
$ pip install lenstronomy --user
\end{lstlisting}
\end{lrbox}

\pythonpackage\ is an open source software package developed in \Python\ \citep{python_1995}. Distribution is granted through the MIT open source software license. \pythonpackage\ relies on packages included in the \Python\ standard library\footnote{\pythonPSL} and the proven open-source libraries \texttt{numpy} \footnote{\url{www.numpy.org}}, \texttt{scipy} \footnote{\url{www.scipy.org}}, \texttt{astropy} \citep{astropy}, and \texttt{matplotlib} \citep{matplotlib}. For optimization and parameter inference routines, \texttt{CosmoHammer} \citep{Akeret:2013nl} is used, which supports Message Passing Interface (MPI) and allows for massive parallelization of the emcee \citep{emcee} algorithm. 

The stable release version can be installed at the command line via pip:\\
\\
\usebox{\LstBox}
\\
\\
\pythonpackage\ is compatible with both \texttt{python2}\footnote{\texttt{python2} will not be maintained past 2020} and \texttt{python3}. The code design and development follow effective practices for scientific computing. The development is coordinated on \texttt{GitHub}\footnote{\pythonpackageGitHub} and stable versions are released through \texttt{PyPi}\footnote{\pythonpackagePyPi} - the python packaging index. This full integration of \pythonpackage\ in the \Python\ packaging environment allows third party to release and share software packages, analysis routines and work-flows that rely upon \pythonpackage.

Extensive tests have been added to \pythonpackage\, which allows us to develop the package using a Continuous Integration (CI) process. The tests are performed on a virtual environment\footnote{\travisLink} to ensure cross-platform stability and the test coverage report is publicly available\footnote{\coverallLink}. Documentation for the various routines and classes are provided through \texttt{sphinx}\footnote{\url{www.sphinx-doc.org}} and are available on \texttt{ReadtheDocs}\footnote{\docLink}. To make use of \pythonpackage, a start guide and several application examples are provided in the form of \texttt{Jupyter} notebooks in an extension module\footnote{\tutorialLink}.

\pythonpackage\ is structured in multiple independent modules, each consisting of multiple classes and sub-packages. These modules execute specific tasks and come with well-defined API. Among the core modules are:

\begin{enumerate}
    \item \texttt{LensModel}: Provides the lensing functionalities. The full functionality is supported with an arbitrary superposition of individual lens models (Section \ref{sec:lens_model}).
    \item \texttt{LightModel}: Enables a variety of surface brightness descriptions and profiles. See Section \ref{sec:light_model}.
    \item \texttt{PointSource}: Handles the point sources (Section \ref{sec:point_source_model}).
    \item \texttt{Data}: Handling all data specific tasks. Including Point-Spread function (PSF), coordinate systems and noise properties (Section \ref{sec:data}).
    \item \texttt{ImSim}: Simulates images. Queries the specifications made in \texttt{LensModel}, \texttt{LightModel}, \texttt{PointSource} and \texttt{Data} (Section \ref{sec:imsim}).
    \item \texttt{Sampling}: Performs the sampling of the parameter space. Aside from the \texttt{Sampling} class that offers pre-defined sampling algorithm, the module includes the , \texttt{Likelihood} class to computes the likelihood based on the \texttt{ImSim} module and the \texttt{Param} class to handle the parameters and their assigned constraints throughout the sampling (Section \ref{sec:sampling}).
    \item \texttt{Workflow}: Higher level API to define fitting sequences and infer model parameters based on the \texttt{Sampling} (Section \ref{sec:workflow}).
    \item \texttt{GalKin}: Computes (stellar) kinematics of the deflector galaxy with spherical Jeans modeling based on the mass model specified in \texttt{LensModel} and the lens light model specified in \texttt{LightModel} (Section \ref{sec:galkin}).
\end{enumerate}

The core modules perform the individual tasks associated with lens modeling. Each module can be used as a stand-alone package and various extension modules are available. The strength of \pythonpackage\ is the full integrated support of each individual module when it comes to lens modeling.

\begin{figure}
  \centering
  \includegraphics[angle=0, width=140mm]{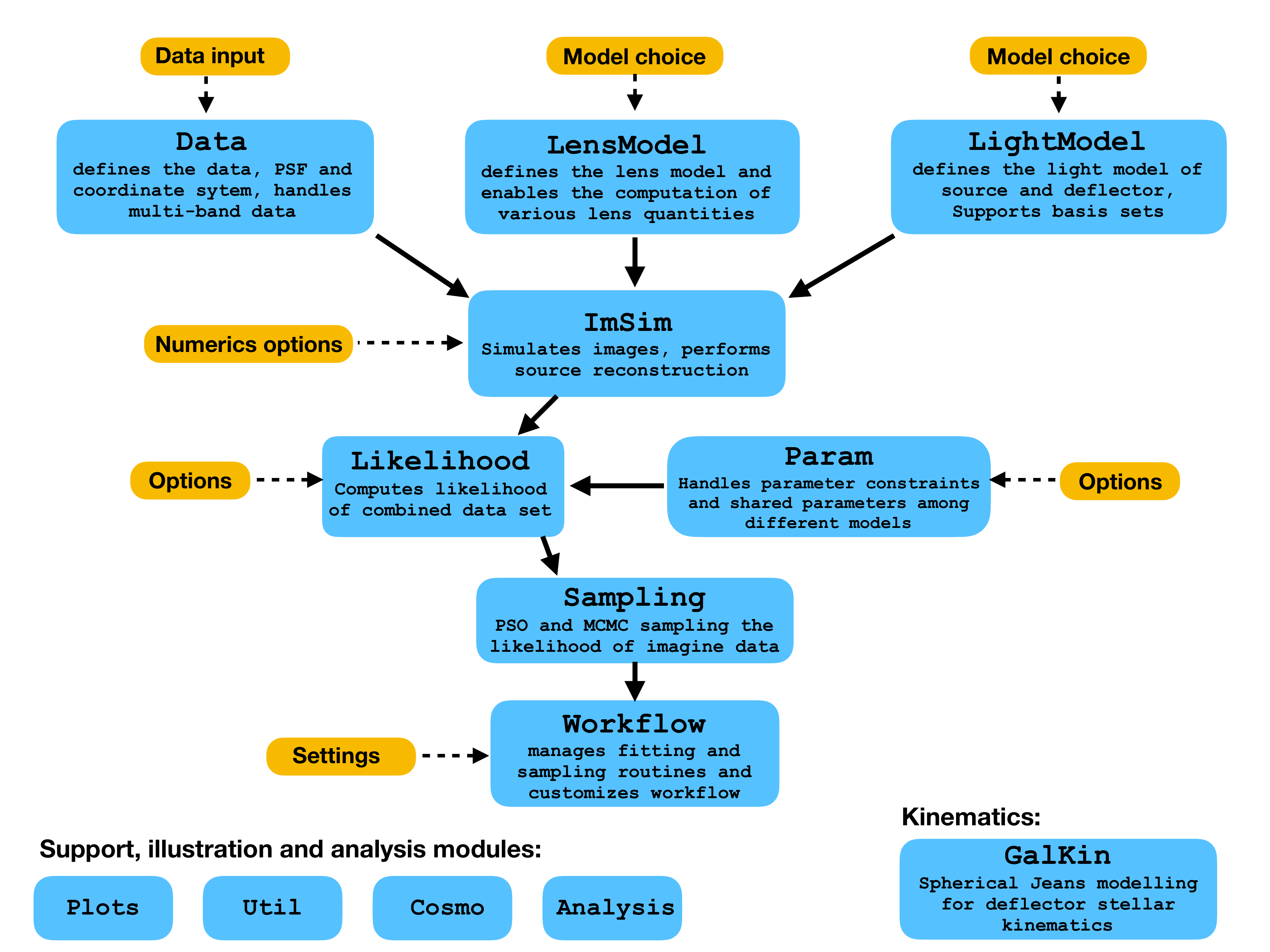}
  \caption{The \pythonpackage\ hierarchical structure of the core modules. Arrow mark the dependencies and interfaces between the different modules. Orange boxes mark user API's.}
\label{fig:flow_chart}
\end{figure}

In Section \ref{sec:core_modules}, we briefly describe the main functionalities of the core modules and provide some simple use cases. The interplay between the modules is demonstrated in Section \ref{sec:model_examples} and \ref{sec:applications}.

\section{Core modules of \pythonpackage}
\label{sec:core_modules}
In the following, we describe the basic functionalities of the most important modules of \pythonpackage\ with some simple examples. More detailed information about the available routines and their use can be accessed through the online documentation.

\subsection{\texttt{LensModel} module}
\label{sec:lens_model}
\texttt{LensModel} and its sub-packages execute all the purely lensing related tasks of \pythonpackage. This includes ray-shooting, solving the lens equation, arrival time computation and non-linear solvers to optimize lens models for specific image configurations. The module allows consistent integration with single and multi plane lensing and an arbitrary superpositions of lens models. There is a wide range of lens models available. For details we refer the reader to the online-documentation.

To demonstrate the design of \texttt{LensModel}, we initialize a lens model and then execute some lensing calculations. First, we perform these calculations in a single-plane configuration \ref{sec:single_plane} and then in a multi-plane configuration \ref{sec:multi_plane}. Then we demonstrate the lens equation solver, that can be applied in both cases with the same API \ref{sec:lens_equation_solver}.

\subsubsection{Single-plane lensing} \label{sec:single_plane}
The default setting of \texttt{LensModel} is to operate in single lens plane mode, where the superpositions of multiple lens models are de-coupled. Below we provide and example of a lens model, that consists of a super-position of an elliptical power-law potential, an external shear and an additional singular isothermal sphere perturber. We initialize the \texttt{LensModel} class, define the parameters for each individual model and perform some standard lensing calculations, such as a backwards ray-shooting of an image plane coordinate, computation of the Fermat potential and evaluating the magnification.

\begin{lstlisting}[language=Python]
# import the LensModel class #
from lenstronomy.LensModel.lens_model import LensModel

# specify the choice of lens models #
lens_model_list = ['SPEP', 'SHEAR', 'SIS']

# setup lens model class with the list of lens models #
lensModel = LensModel(lens_model_list=lens_model_list)

# define parameter values of lens models #
kwargs_spep = {'theta_E': .9, 'e1': 0.05, 'e2': 0.05, 'gamma': 2., 'center_x': 0.1, 'center_y': 0}
kwargs_shear = {'e1': 0.05, 'e2': 0.0}
kwargs_sis = {'theta_E': 0.1, 'center_x': 1., 'center_y': -0.1}
kwargs_lens = [kwargs_spep, kwargs_shear, kwargs_sis]

# image plane coordinate #
theta_ra, theta_dec = .9, .4

# source plane coordinate #
beta_ra, beta_dec = lensModel.ray_shooting(theta_ra, theta_dec, kwargs_lens)
# Fermat potential #
fermat_pot = lensModel.fermat_potential(x_image=theta_ra, y_image=theta_dec, x_source=beta_ra, y_source=beta_dec, kwargs_lens=kwargs_lens)

# Magnification #
mag = lensModel.magnification(theta_ra, theta_dec, kwargs_lens)
\end{lstlisting}

Additionally, the \texttt{LensModel} class allows to compute the Hessian matrix, shear and convergence, deflection angle and lensing potential. These routines are fully compatible with the \texttt{numpy} array structure and superposition of an arbitrary number of lens models.

\subsubsection{Multi-plane lensing} \label{sec:multi_plane}
The multi-plane setting of \texttt{LensModel} allows the user to place several deflectors at different redshifts. When not further specified, the default cosmology used is that of the \texttt{astropy} cosmology class. The API to access the lensing functionalities remains the same as for the single-plane setting \ref{sec:single_plane}. As an example, we take the same setting as in \ref{sec:single_plane} but place the singular isothermal sphere perturber at a lower redshift.
\begin{lstlisting}[language=Python]
# keep the imports and variables from above #
# specify redshifts of deflectors #
redshift_list = [0.5, 0.5,.1]
# specify source redshift #
z_source = 1.5
# setup lens model class with the list of lens models #
lensModel_mp = LensModel(lens_model_list=lens_model_list, z_source=z_source, redshift_list=redshift_list, multi_plane=True)

# source plane coordinate #
beta_ra, beta_dec = lensModel_mp.ray_shooting(theta_ra, theta_dec, kwargs_lens)

# Magnification #
mag = lensModel_mp.magnification(theta_ra, theta_dec, kwargs_lens)
\end{lstlisting}

\subsubsection{Lens equation solver} \label{sec:lens_equation_solver}
Solving the lens equation to compute the (multiple) image positions of a given source position can be conveniently performed within \texttt{LensModel} and is supported with a general instance of the \texttt{LensModel} class.
\begin{lstlisting}[language=Python]
# keep the imports and variables from above #
# import the lens equation solver class #
from lenstronomy.LensModel.Solver.lens_equation_solver import LensEquationSolver

# specifiy the lens model class to deal with #
solver = LensEquationSolver(lensModel)

# solve for image positions provided a lens model and the source position #
theta_ra, theta_dec = solver.image_position_from_source(beta_ra, beta_dec, kwargs_lens)

# the magnification of the point source images #
mag = lensModel.magnification(theta_ra, theta_dec, kwargs_lens)
\end{lstlisting}

\begin{figure}
  \centering
  \includegraphics[angle=0, width=140mm]{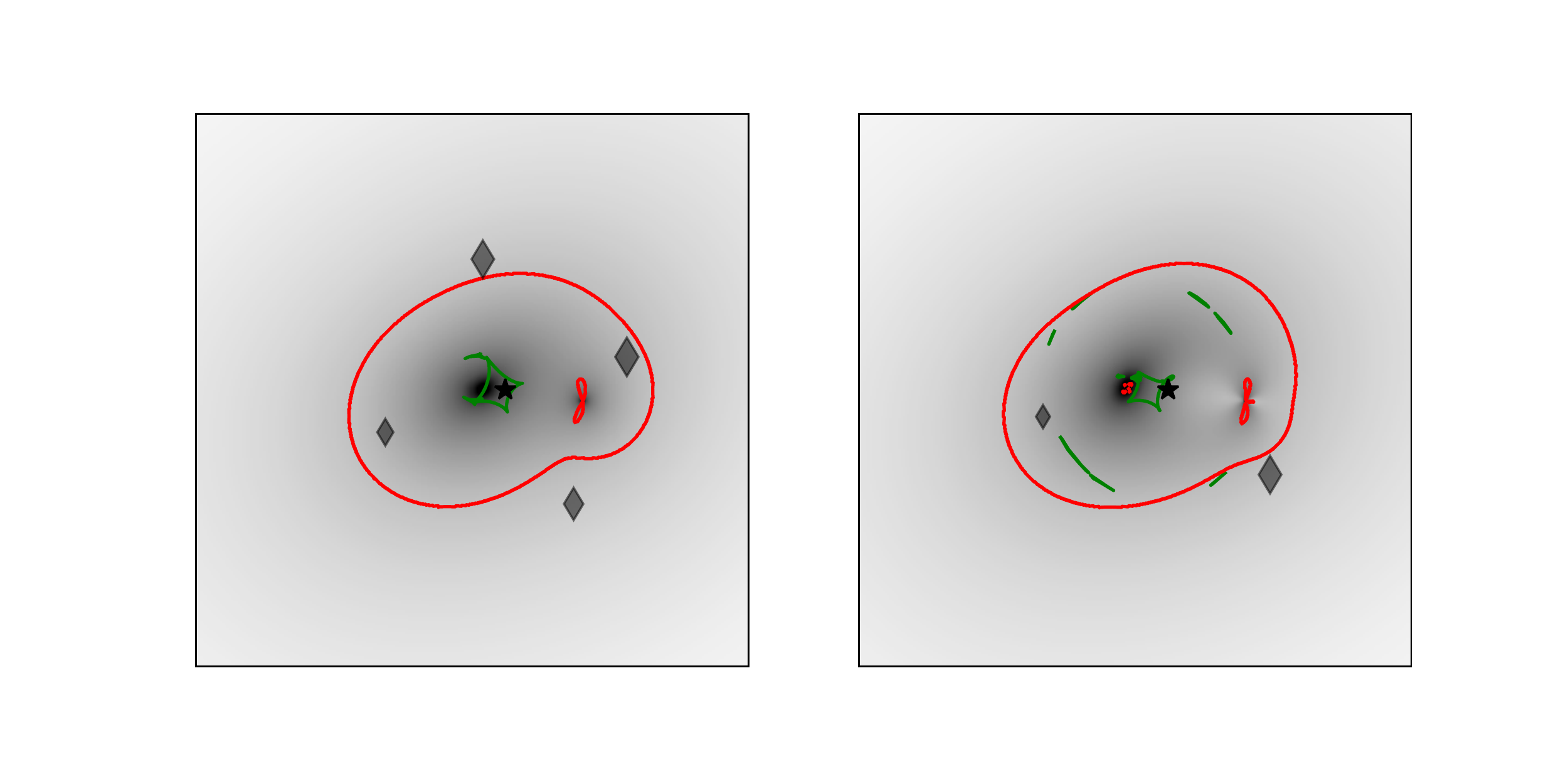}
  \caption{Illustrations of the two lens models created in Section \ref{sec:lens_model}. \textbf{Left:} The single-plane lens model of \ref{sec:single_plane}. \textbf{Right:} The multi-plane lens model of \ref{sec:multi_plane}. In gray-scale the convergence map of the lens models, red lines correspond to the critical curve, green lines to the caustics. The green star corresponds to the source positions and the diamonds to the corresponding image positions.}
\label{fig:lens_model}
\end{figure}

Two lens models are shown in Figure \ref{fig:lens_model}. The source position of the example and the solutions of the lens equation (image positions) are marked.

\subsection{\texttt{LightModel} module}
\label{sec:light_model}
The \texttt{LightModel} class provides the functionality to describe galaxy surface brightnesses. \texttt{LightModel} supports various analytic profiles as well as representations in shapelet basis sets. Any superposition of different profiles is supported. We refer to the online documentation for the full list of surface brightness profiles available and their parameterisation.

As an example, we initialize two \texttt{LightModel} class, one with a spherical Sersic profile and one with an elliptical Sersic profile. We define the profile parameters and evaluate the surface brightness at a specific position. The two \texttt{LightModel} instances will later be used as the lens light and the source light.

\begin{lstlisting}[language=Python]
# import the LightModel class #
from lenstronomy.LightModel.light_model import LightModel
# set up the list of light models to be used #
source_light_model_list = ['SERSIC']
lightModel_source = LightModel(light_model_list=source_light_model_list)
lens_light_model_list = ['SERSIC_ELLIPSE']
lightModel_lens = LightModel(light_model_list=lens_light_model_list)
# define the parameters #
kwargs_light_source = [{'I0_sersic': 10, 'R_sersic': 0.02, 'n_sersic': 1.5, 'center_x': beta_ra, 'center_y': beta_dec}]
# ellipticity parameter are defined as eccentricities #
import lenstronomy.Util.param_util as param_util
e1, e2 = param_util.phi_q2_ellipticity(phi=0.5, q=0.7)
kwargs_light_lens = [{'I0_sersic': 1000, 'R_sersic': 0.1, 'n_sersic': 2.5, 'e1': e1, 'e2': e2, 'center_x': 0.1, 'center_y': 0}]

# evaluate surface brightness at a specific position #
flux = lightModel_lens.surface_brightness(x=1, y=1, kwargs_list=kwargs_light_lens)
\end{lstlisting}

\subsection{\texttt{PointSource} module}
\label{sec:point_source_model}
To accurately predict and model the positions and fluxes of point sources, different numerical procedures are needed compared to extended surface brightness features. The \texttt{PointSource} module manages the different options in describing point sources (e.g. in the image plane or source plane, with fixed magnification or allowed with individual variations thereof) and provides a homogeneous API to access image positions and magnifications. The \texttt{PointSource} class requires an instance of a \texttt{LensModel} class in case of lensed sources and arbitrary superpositions of point sources are allowed.

In the example below, we create two instances of the \texttt{PointSource} class. One with a parameterization in the source plane and one with a parameterization in the image plane. The API to access the necessary information about the image positions and magnifications remain the same in both cases.

\begin{lstlisting}[language=Python]
# import the PointSource class #
from lenstronomy.PointSource.point_source import PointSource

# unlensed source positon #
point_source_model_list = ['SOURCE_POSITION']
pointSource = PointSource(point_source_type_list=point_source_model_list, lensModel=lensModel, fixed_magnification_list=[True])
kwargs_ps = [{'ra_source': beta_ra, 'dec_source': beta_dec, 'source_amp': 100}]
# return image positions and amplitudes #
x_pos, y_pos = pointSource.image_position(kwargs_ps=kwargs_ps, kwargs_lens=kwargs_lens)
point_amp = pointSource.image_amplitude(kwargs_ps=kwargs_ps, kwargs_lens=kwargs_lens)

# lensed image positions (solution of the lens equation) #
point_source_model_list = ['LENSED_POSITION']
pointSource = PointSource(point_source_type_list=point_source_model_list, lensModel=lensModel, fixed_magnification_list=[False])
kwargs_ps = [{'ra_image': theta_ra, 'dec_image': theta_dec, 'point_amp': np.abs(mag)*30}]
# return image positions and amplitudes #
x_pos, y_pos = pointSource.image_position(kwargs_ps=kwargs_ps, kwargs_lens=kwargs_lens)
point_amp = pointSource.image_amplitude(kwargs_ps=kwargs_ps, kwargs_lens=kwargs_lens)
\end{lstlisting}

\subsection{\texttt{Data} module}
\label{sec:data}
The \texttt{Data} module consists of two main classes. The \texttt{Data} class stores and manages all the imaging data relevant information. This includes the coordinate frame, coordinate-to-pixel transformation (and the inverse), and, in the case of fitting, also noise properties for computing the likelihood of the data given the model. The \texttt{PSF} class handles the point spread function convolution. Supported are pixelised convolution kernels as well as some analytic profiles.
\begin{lstlisting}[language=Python]
# import the Data() class #
from lenstronomy.Data.imaging_data import Data
deltaPix = 0.05  # size of pixel in angular coordinates #

# setup the keyword arguments to create the Data() class #
kwargs_data = {'numPix': 100,
    'ra_at_xy_0': -2.5,
    'dec_at_xy_0': -2.5, 
    'transform_pix2angle': np.array([[1, 0], [0, 1]]) * deltaPix}
data = Data(kwargs_data)
# return the list of pixel coordinates #
x_coords, y_coords = data.coordinates
# compute pixel value of a coordinate position #
x_pos, y_pos = data.map_coord2pix(ra=0, dec=0)
# compute the coordinate value of a pixel position #
ra_pos, dec_pos = data.map_pix2coord(x=20, y=10)

# import the PSF() class #
from lenstronomy.Data.psf import PSF
kwargs_psf = {'psf_type': 'GAUSSIAN', 'fwhm': 0.1, 'pixel_size': deltaPix}
psf = PSF(kwargs_psf)
# return the pixel kernel corresponding to a point source # 
kernel = psf.kernel_point_source
\end{lstlisting}

\subsection{\texttt{ImSim} module}
\label{sec:imsim}
At the core of the \texttt{IMSim} module is the \texttt{ImageModel} class. \texttt{ImageModel} is the interface to combine all the different components, \texttt{LensModel}, \texttt{LightModel}, \texttt{PointSource} and \texttt{Data} to model images. The \texttt{LightModel} can be used to model both lens light (un-lensed) and source light (lensed) components. \texttt{ImSim} supports all functionalities of each of those components. \texttt{ImageModel} is supported by the class \texttt{ImageNumerics} that specifies and executes the numerical options accessible. Among the numerical options are sub-pixel grid resolution ray-tracing and convolutions that can improve numerical accuracy in the presence of either small lensing perturbations and/or a highly variable surface brightness profile \citep[see e.g.][for the latter]{Tessore:2016}.

\subsubsection{Image simulation} \label{sec:image_simulation}
As an example, we simulate an image with an instance of \texttt{ImageModel} that use instances of the classes we created above. We can define two different \texttt{LightModel} instances for the lens and source light. We define the sub-pixel ray-tracing resolution and whether the PSF convolution is applied on the higher resolution ray-tracing grid or on the degraded pixel image.
\begin{lstlisting}[language=Python]
# import the ImageModel class #
from lenstronomy.ImSim.image_model import ImageModel
# define the numerics #
kwargs_numerics = {'subgrid_res': 3,
                  'psf_subgrid': True}
# initialize the Image model class by combining the modules we created above #
imageModel = ImageModel(data_class=data, psf_class=psf, lens_model_class=lensModel,
                        source_model_class=lightModel_source,
                        lens_light_model_class=lightModel_lens,
                        point_source_class=pointSource, 
                        kwargs_numerics=kwargs_numerics)
# simulate image with the parameters we have defined above #
image = imageModel.image(kwargs_lens=kwargs_lens, kwargs_source=kwargs_light_source,
                         kwargs_lens_light=kwargs_light_lens, kwargs_ps=kwargs_ps)

# we can also add noise #
import lenstronomy.Util.image_util as image_util
# exposure time to quantify the Poisson noise level #
exp_time = 100
# background rms value #
background_rms = 0.1  
poisson = image_util.add_poisson(image, exp_time=exp_time)
bkg = image_util.add_background(image, sigma_bkd=background_rms)
image_noisy = image + bkg + poisson
\end{lstlisting}

\begin{figure}
  \centering
  \includegraphics[angle=0, width=140mm]{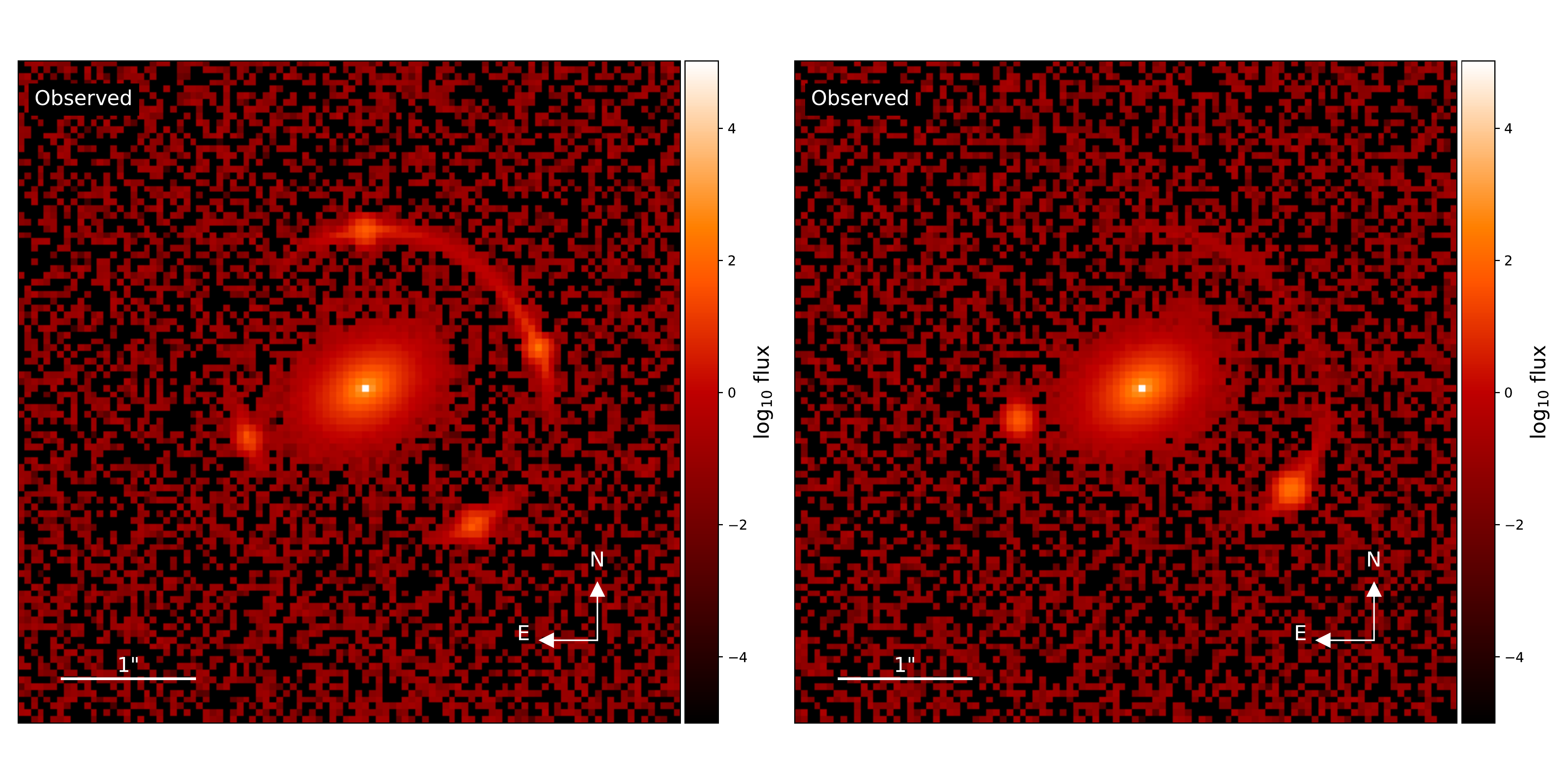}
  \caption{Simulated images with the chosen options of the \texttt{ImageModel} class (\ref{sec:imsim}). \textbf{Left:} The single-plane lens model results in a quadruply imaged quasar. \textbf{Right:} The multi-plane model results in a double lensed quasar.}
\label{fig:example_quad}
\end{figure}

\begin{figure*}
  \centering
  \includegraphics[angle=0, width=140mm]{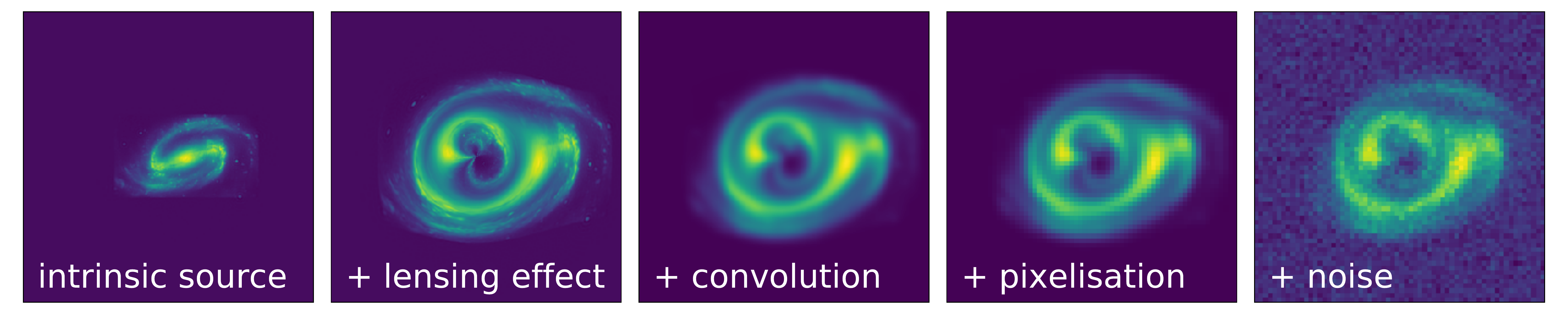}
  \caption{Illustration of the sequence of steps performed by the \texttt{ImSim} module. \textbf{From left to right:} A galaxy described with a high resolution shapelet representation (1), seen through gravitational lensing, i.e. evaluated in the image plane coordinates (2), convolved on a subpixel high resolution grid (3), down-graded to pixel resolution and added with noise (5).}
\label{fig:computation_illustration}
\end{figure*}

Figure \ref{fig:example_quad} shows the simulated image of the example computed above with the single-plane lens model of Section \ref{sec:single_plane} (left panel) and for the same Sersic light profiles but with the multi-plane lens model of Section \ref{sec:multi_plane} in the right panel. To illustrate the numerical procedure in how \pythonpackage\ renders images, we provide another example consisting of a high resolution galaxy profile in Figure \ref{fig:computation_illustration}.

\subsubsection{Linear inversion} \label{sec:linear_inversion}
Parameters corresponding to an an amplitude of a surface brightness distribution have a linear response on the predicted flux values of pixels and can be inferred by a linear minimization based on the data \citep{Warren:2003eg}. \pythonpackage\ automatically identifies those parameters. The \texttt{ImSim} module comes with an option such that the linear parameters do not have to be provided when fitting a model to data. This can reduce the number of non-linear parameters significantly, depending on the source complexity to be modelled. In the example provided in Section \ref{sec:image_simulation}, we have 6 linear parameters, the 4 point source amplitudes and the amplitudes of the Sersic profile of the lens and source. To perform the linear inversion, noise properties of the data have to be known or assumed (see Section \ref{sec:likelihood}). There are different approaches in the literature that perform different types of semi-linear inversions \citep[e.g.][]{Suyu:2006, Vegetti:2009, Tagore:2014, Birrer2015_basis_set, Nightingale:2015}.

In the example below, we add the noisy data to the \texttt{ImageModel} instance, then delete the knowledge about the linear parameters and solve for the linear coefficients based on the data.
\begin{lstlisting}[language=Python]
# update the data with the noisy image and its noise properties #
kwargs_data['image_data'] = image_noisy
kwargs_data['background_rms'] = background_rms
kwargs_data['exposure_map'] = np.ones_like(image_noisy)* exp_time
data_class_sim = Data(kwargs_data)
imageModel.update_data(data_class_sim)

# we do not require the knowledge of the linear parameters #
del kwargs_light_source[0]['amp']
# apply the linear inversion to fit for the noisy image #
image_reconstructed, _, _, _ = imageModel.image_linear_solve(kwargs_lens=kwargs_lens, kwargs_source=kwargs_light_source, kwargs_lens_light=kwargs_light_lens, kwargs_ps=kwargs_ps)
\end{lstlisting}

\subsubsection{Likelihood definition} \label{sec:likelihood}
The likelihood of the data given a model $p(d_{\text{data}}|d_{\text{model}})$ is key in sampling the parameter posterior distribution (Section \ref{sec:sampling}) and also to perform the linear inversion (Section \ref{sec:linear_inversion}). The convention \pythonpackage\ uses to compute $p(d_{\text{data}}|d_{\text{model}})$ is
\begin{equation} \label{eqn:likelihood}
 \log p(d_{\text{data}}|d_{\text{model}}) = \sum_i \frac{(d_{\text{data,i}} - d_{\text{model,i}})^2}{2\sigma_{\text{i}}^2} + \text{const}.
\end{equation}
The constant term in equation \ref{eqn:likelihood} is not computed by \pythonpackage.
The error in each pixel, $\sigma_{\text{i}}$, consists of a Gaussian background term, $\sigma_{\text{bkg}}$, and a Poisson term based on the count statistics of an individual pixel, $f_i$, such that $d_{\text{model,i}} / f_i$ is the Poisson error predicted by the model in the time units of the data, and writes
\begin{equation}
    \sigma_{i}^2 = \sigma_{\text{bkgd}}^2 + d_{\text{model,i}}/f_i.
\end{equation}
In our example of \ref{sec:image_simulation}, $f_i$ is the exposure time for each pixel and $\sigma_{\text{bkgd}}$ is the background rms value. CCD gain and other components may be incorporated into $f_i$.

The linear inversion requires an estimate of the noise term, $\sigma_{i}^2$, without the knowledge of the model, $d_{\text{model,i}}$. For this particular step, the linear inversion is performed based on the Poisson noise expected by the data itself
\begin{equation}
    \sigma_{\text{linear, i}}^2 = \sigma_{\text{bkgd}}^2 + d_{\text{data,i}}/f_i.
\end{equation}

The analytic marginalization over the covariance matrix of the linear inversion (Gaussian approximation) can be added \citep[see][for further information]{Birrer2015_basis_set}. Additionally, pixel masks can be set and additional error terms can be plugged in, if required. \pythonpackage\ provides a direct access to the likelihood of the data given a model and performs all the required computations:
\begin{lstlisting}[language=Python]
logL = imageModel.likelihood_data_given_model(kwargs_lens,kwargs_light_source,kwargs_light_lens,kwargs_ps)
\end{lstlisting}

\subsection{\texttt{Sampling} module} \label{sec:sampling}
The \texttt{Sampling} module manages the execution of the non-linear fitter (e.g. PSO) and the parameter inference (e.g. emcee). The module is built up such that the user can plug in their own customized sampler.
The \texttt{Sampling} Module consists of three major classes: The \texttt{Likelihood} class manages the specific likelihood function, consisting of the imaging likelihood and potential other data and constraints and provides the interface to the sampling routines. The \texttt{Param} class handles all the model choices and the parameters going in it and supports the \texttt{Likelihood} class. Together they handle all the model choices of the user and mitigate them to the external modules and from the external modules back to \pythonpackage.
Finally, the \texttt{Sampler} class gives specific examples how the \texttt{Likelihood} class can be used to execute specific samplers.


\subsubsection{Parameter handling}
External sampling modules require a likelihood function that is consistent with their own parameter handling, mostly in ordered arrays. The likelihood in Section \ref{sec:likelihood} requires \pythonpackage\ conventions in terms of lists of keyword arguments.
The \texttt{Param} class is the API of the \pythonpackage\ conventions of parameters used in the \texttt{ImSim} module and the standardized parameter arrays used by external samplers (such as \texttt{CosmoHammer} or \texttt{emcee}). The \texttt{Param} class enables the user further to set options:
\begin{enumerate}
    \item keep certain parameters fixed
    \item handling of the linear parameters
    \item provide additional constraints on the modelling (e.g. fix source profile to point source position etc.)
\end{enumerate}
Below we provide an example where initialize a \texttt{Param} class consistent with the options chosen in the previous sections and where we specify fixed and joint parameters. We then perform the mapping between \pythonpackage\ conventions and formats being used by external sampling modules.
\begin{lstlisting}[language=Python]
# import the parameter handling class #
from lenstronomy.Sampling.parameters import Param
# set options for constraint parameters #
kwargs_constraints = {
                'num_point_source_list': [4],  # when modelling a doube, use [2]
                'joint_source_with_point_source': [[0, 0]],
                'solver': True,
                'solver_type': 'PROFILE_SHEAR',  # 'PROFILE', 'PROFILE_SHEAR', 'ELLIPSE', 'CENTER'
               }
# list fixed parameters #
kwargs_fixed_lens = [{'gamma': 2.}, {}, {}]
kwargs_fixed_source = [{}]
kwargs_fixed_lens_light = [{}]
kwargs_fixed_ps = [{}]

# make instance of parameter class with given model options, constraints and fixed parameters #
param = Param(kwargs_model, kwargs_constraints, kwargs_fixed_lens, kwargs_fixed_source,
              kwargs_fixed_lens_light, kwargs_fixed_ps, kwargs_lens_init=kwargs_lens)

# the number of non-linear parameters and their names #
num_param, param_list = param.num_param()

# parameter array for fitting routine #
param_array = param.setParams(kwargs_lens, kwargs_light_source, kwargs_light_lens, kwargs_ps)

# recover keyword arguments list for lenstronomy from a parameter array of the fitting routine #
kwargs_lens_out, kwargs_light_source_out, kwargs_light_lens_out, kwargs_ps_out, _ = param.getParams(param_array)

# print settings #
param.print_setting()
\end{lstlisting}

\subsubsection{Likelihood execution}
The \texttt{Likelihood} class combines the \texttt{ImSim} module and the \texttt{Param} class to allow a direct access to the \pythonpackage\ likelihood from an external sampler. In addition, the \texttt{Likelihood} class allows to a simultaneous handling of multi-band data and to incorporate other data, such as time-delay measurements. Below we initialize a \texttt{Likelihood} class and execute the likelihood function from an ordered array of parameters.
\begin{lstlisting}[language=Python]
# define options of the likelihood #
kwargs_likelihood = {'source_marg': False,
                    'check_solver': True,
                    'solver_tolerance': 0.001,}

# import the likelihood class #
from lenstronomy.Sampling.likelihood import LikelihoodModule

# make instance of the likelihood class #
likelihoodModule = LikelihoodModule(imSim_class=imageModel, param_class=param, kwargs_likelihood=kwargs_likelihood)

# execute the likelihood from an ordered parameter array #
logL = likelihoodModule.logL(args=param_array)
\end{lstlisting}

\subsubsection{Sampling the parameter space}
The \texttt{Sampler} class consists of examples of different samplers that can be used. As an example, we run a Particle Swarm Optimization (PSO) with the previous instance of the \texttt{Likelihood} class.
\begin{lstlisting}[language=Python]
# initial guess of non-linear parameters, we chose different starting parameters than the truth #
kwargs_lens_init = [{'theta_E': 1.2, 'e1': 0, 'e2': 0, 'gamma': 2., 'center_x': 0., 'center_y': 0},
    {'e1': 0.0, 'e2': 0.01}, {'theta_E': 0.1, 'center_x': 1., 'center_y': 0}]
kwargs_source_init = [{'R_sersic': 0.03, 'n_sersic': 1., 'center_x': 0, 'center_y': 0}]
kwargs_lens_light_init = [{'R_sersic': 0.1, 'n_sersic': 1, 'e1': 0, 'e2': 0, 'center_x': 0, 'center_y': 0}]
kwargs_ps_init = [{'ra_image': theta_ra+0.01, 'dec_image': theta_dec-0.01}]

# initial spread in parameter estimation #
kwargs_lens_sigma = [{'theta_E': 0.3, 'e1': 0.5, 'e2': 0.5, 'gamma': .2, 'center_x': 0.1, 'center_y': 0.1},
    {'e1': 0.1, 'e2': 0.1}, {'theta_E': 0.1, 'center_x': .1, 'center_y': 0.1}]
kwargs_source_sigma = [{'R_sersic': 0.1, 'n_sersic': .5, 'center_x': .1, 'center_y': 0.1}]
kwargs_lens_light_sigma = [{'R_sersic': 0.1, 'n_sersic': 0.2, 'e1': 0.2, 'e2': 0.2, 'center_x': .1, 'center_y': 0.1}]
kwargs_ps_sigma = [{'ra_image': [0.02] * 4, 'dec_image': [0.02] * 4}]

# ordered array for sampler #
param_init = param.setParams(kwargs_lens_init, kwargs_source_init, kwargs_lens_light_init, kwargs_ps_init)
param_sigma = param.setParams(kwargs_lens_sigma, kwargs_source_sigma, kwargs_lens_light_sigma, kwargs_ps_sigma)

# upper and lower range to initialise the sampler #
upper_start = np.array(param_init) + np.array(param_sigma)
lower_start = np.array(param_init) - np.array(param_sigma)

# import the sampling class #
from lenstronomy.Sampling.sampler import Sampler
# initialize the Sampler class with the likelihoodModule #
sampler = Sampler(likelihoodModule=likelihoodModule)

# execute a PSO from the Sampler #
result, chain_properties = sampler.pso(n_particles=200, n_iterations=200, lower_start=lower_start, upper_start=upper_start)

# cast result back in lenstronomy conventions #
lens_result, source_result, lens_light_result, ps_result, _ = param.getParams(result)
\end{lstlisting}
Additionally to the example mentioned above, hard bounds on the upper and lower range in parameter space can be provided.

\subsection{\texttt{Workflow} module} \label{sec:workflow}

The \texttt{Workflow} module allows the user to perform a sequence of PSO and/or MCMC runs. The user can run the sequence of fitting routines with taking the results of the previous routine as an input of the next one. The user can specify (optionally) to keep one or multiple parameter classes (lens model, source model, lens light model and source model) fixed during the fitting process of individual runs. Iterative PSF optimization can also be injected within the fitting sequence.
The \texttt{FittingSequence} class enables a reliable execution of tasks on non-local platforms, such as hight performance clusters and supports parallel executions of likelihood evaluations with the MPI portocoll built in \texttt{CosmoHammer} \citep{Akeret:2013nl}.

\subsection{\texttt{GalKin} module}
\label{sec:galkin}
Kinematics of the lensing galaxy can provide additional constraints on the lens model and can help to reduce systematics inherent in lensing. The \texttt{GalKin} module provides the support to self-consistently model and predict the velocity dispersion of the lensing galaxy given the surface brightness profile and the lens model upon which the image modelling consists of. The kinematics require the knowledge/assumption of the 3d light and mass profiles. Not all lens and light models can be analytically de-projected. In these cases, \pythonpackage\ performs a Multi-Gaussian decomposition \citep{Cappellari:2002_mge} and the de-projection is performed on the individual Gaussian components. The kinematics is computed with spherical Jeans anisotropy modelling (JAM). \pythonpackage\ supports the stellar anisotropy profiles described in \cite{MamonLokas:2005}. Observational conditions, i.e. the PSF and the aperture are modelled with a spectral rendering approach described in \cite{Birrer:2017a}.

\section{Modelling examples} \label{sec:model_examples}
The design of \pythonpackage\ and the core modules described in Section \ref{sec:core_modules} allow a wide range of modelling tasks to be executed. We demonstrated in the previous section how to combine the modules to enable a joint sampling of point source, extended source, lens light and lens deflector model. In this section, we provide five examples in different sub-domains where we demonstrate the capabilities of \pythonpackage, source reconstruction \ref{sec:source_reconstruction}, image de-convolution \ref{sec:deconvolution}, galaxy structural analysis \ref{sec:galfitting}, quasar-host galaxy decomposition \ref{sec:qso_host_decomp} and multiband fitting \ref{sec:multi_band_fitting}. Detailed example workflows for the different applications are presented in the online documentation.

\subsection{Source reconstruction} \label{sec:source_reconstruction}
\begin{figure}
  \centering
  \includegraphics[angle=0, width=140mm]{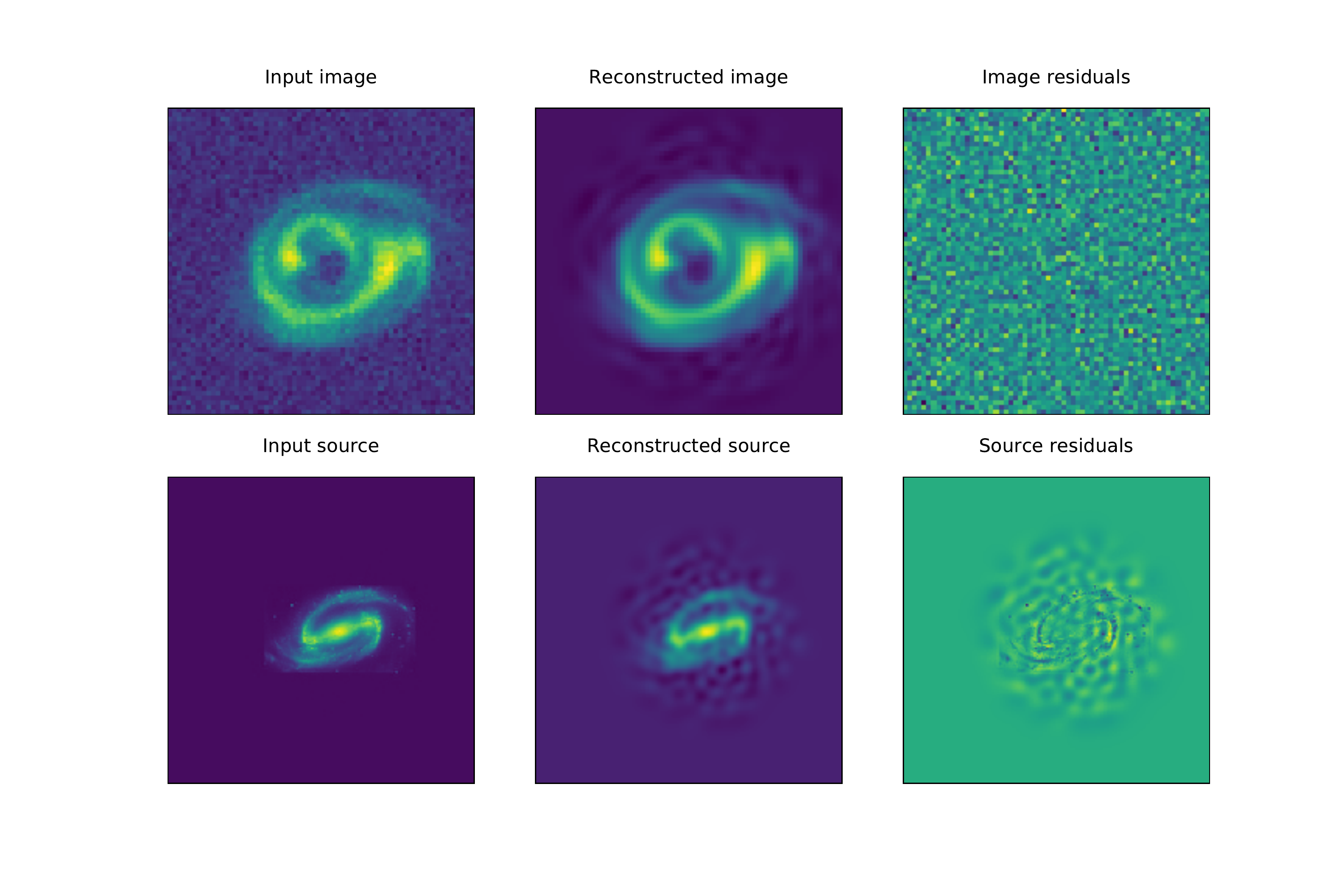}
  \caption{\pythonpackage\ source reconstruction capabilities (de-lensing + de-convolution). \textbf{Top:} Image plane. \textbf{bottom:} Source plane. \textbf{From left to right:} The simulated image with noise based on a realistic galaxy surface brightness profile (same example as Figure \ref{fig:computation_illustration}) (1), the reconstruction by the linear inversion technique based on a lower resolution shapelet basis set (2), the residuals of the input relative to the reconstructed model. The lens model was kept fixed to the input model for the reconstruction.}
\label{fig:de_lens}
\end{figure}

Reconstruction techniques are required to describe the source morphology at the scales relevant for given data. The needed complexity may strongly depend on the type of galaxy being lensed and the resolution and signal-to-noise of the data. In Figure \ref{fig:de_lens}, we provide an example where we reconstruct a source galaxy with complex morphology with a Shapelet basis set with maximum polynomial order, $n_{\text{max}} = 29$. We are able to represent the features present in the image. The reconstruction of the source reproduces the macroscopic morphology of the input galaxy.

\pythonpackage supports a wide range in models and also allows to superpose analytical models with basis sets \citep[see e.g.][]{Birrer:J1206}. The reconstruction for a given set of lens and light model parameters is performed by the linear lens inversion (Section \ref{sec:linear_inversion}). \pythonpackage\ does not provide a Bayesian evidence optimization itself, but this can be performed by the user in post-processing \citep[e.g.][]{Birrer:J1206}.
The performance of the source reconstruction capabilities has been compared with the \texttt{SLIT} software \citep{Joseph:2018} and was found to behave well in speed and reconstruction accuracy.

\subsection{Image de-convolution} \label{sec:deconvolution}
\begin{figure}
  \centering
  \includegraphics[angle=0, width=140mm]{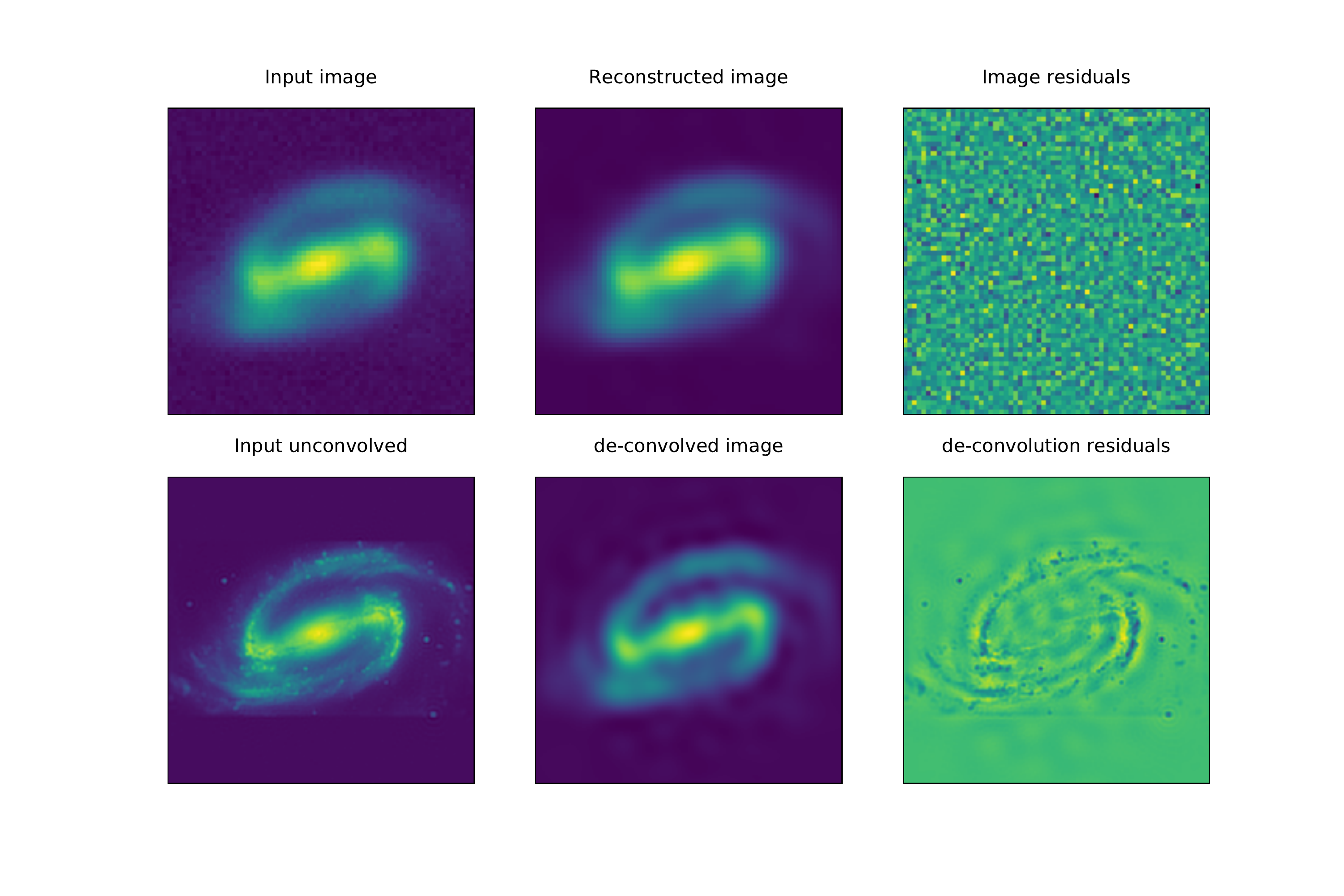}
  \caption{\pythonpackage\ de-convolution capabilities (source reconstruction without lensing). \textbf{Top:} Convolved images. \textbf{bottom:} Unconvolved images. \textbf{From left to right:} The simulated image with noise based on a realistic galaxy surface brightness profile (same example as Figure \ref{fig:computation_illustration} but scaled and with a larger convolution kernel applied) (1), the reconstruction by the linear inversion technique based on a lower resolution shapelet basis set (2), the residuals of the input relative to the reconstructed model.}
\label{fig:de_convolve}
\end{figure}

The source reconstruction in Section \ref{sec:source_reconstruction} is a combination of two distinct steps: a de-lensing (effectively a non-linear mapping between the image plane and the basis set represented in the source plane) and a de-convolution. By removing the class instance of \texttt{LensModel} from the \texttt{ImSim} module or by removing all the lens models, the linear inversion method built in \pythonpackage\ effectively performs a de-convolution. This is demonstrated in Figure \ref{fig:de_convolve} where we take a scaled version of the same galaxy as for Figure \ref{fig:de_lens} with a PSF convolution kernel and apply the same shapelet basis set to describe the image.

\subsection{Galaxy structural analysis} \label{sec:galfitting}
\pythonpackage\ can be used to extract structural components from galaxy images. This is yet another example where the lensing capabilities of \pythonpackage\ do not have to be used necessarily. In terms of flexibility, \pythonpackage\ contains similar features as the well established software \texttt{GALFIT} \citep{Peng:2002_galfit, Peng:2010_galfit}. \pythonpackage\ provides an open source alternative in \Python. We also emphasize that \pythonpackage\ comes along with an MCMC algorithm that can provide covariances between inferred parameters. Additionally, \pythonpackage\ is able to extract structural parameters from lensed and highly distorted galaxies.

\subsection{Quasar-host galaxy decomposition} \label{sec:qso_host_decomp}
\begin{figure}
  \centering
  \includegraphics[angle=0, width=140mm]{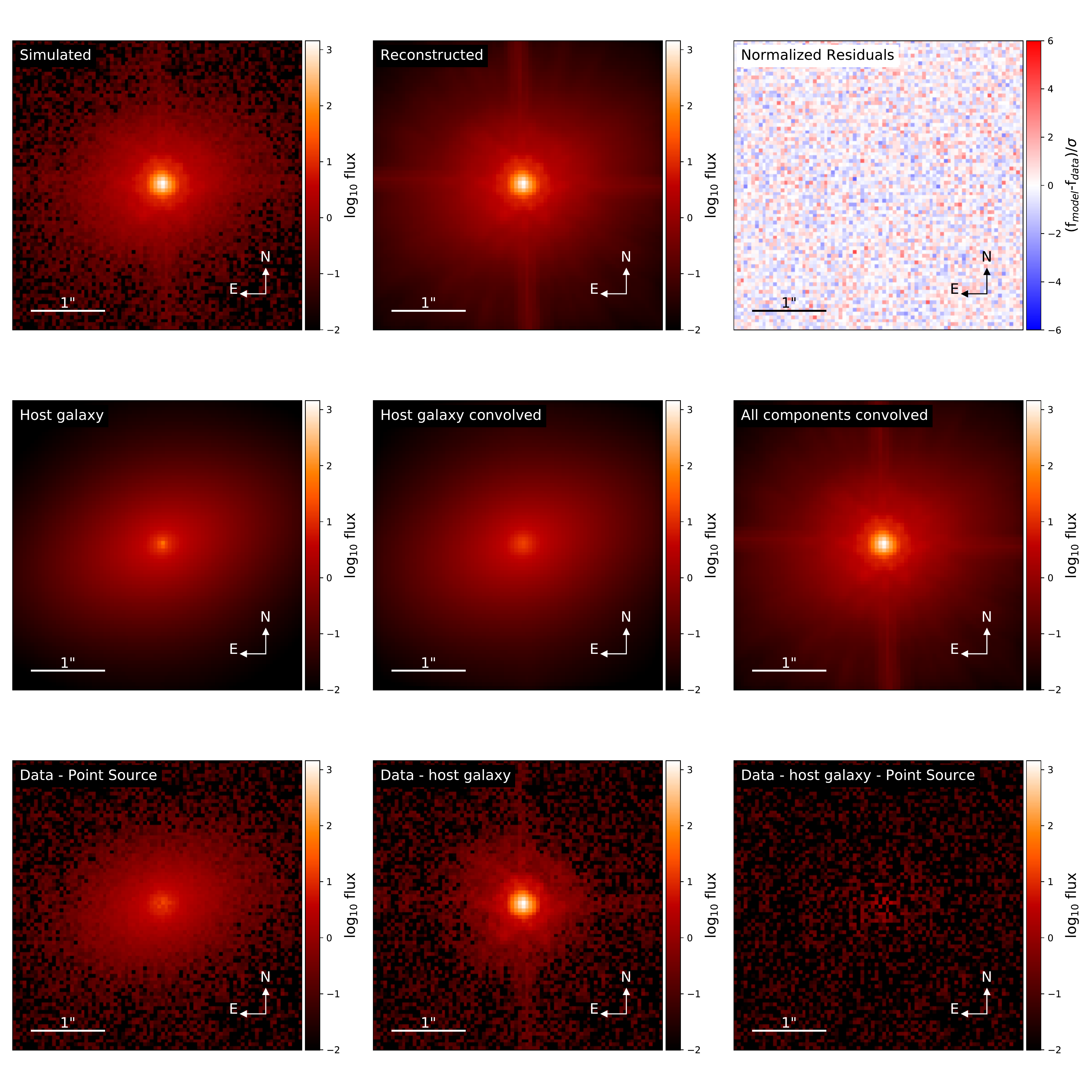}
  \caption{Illustration of the \pythonpackage\ quasar-host galaxy decomposition capabilities. A two-component host galaxy is modelled with a significantly brighter quasar at its center. \pythonpackage\ can reliably decompose the two components. \textbf{Top:} Simulated image (left), reconstructed model (middle) and residuals of the reconstruction (right). \textbf{Middle:} The separation of the model components. \textbf{Bottom:} The subtraction of the model components from the data.}
\label{fig:quasar_host_decomposition}
\end{figure}

In the case where the galaxy contains a quasar, simultaneous decompositions of the host galaxy and a point source component can be performed with \pythonpackage. Figure \ref{fig:quasar_host_decomposition} demonstrates this capability. A joint fitting of two component Sersic profile for the host galaxy and a quasar point source were used as an input model and the different components were recovered in the modelling.

\subsection{Multiband fitting} \label{sec:multi_band_fitting}
\pythonpackage\ is explicitly designed to simultaneously model lenses in multiple imaging bands. The coordinate system definition is image independent and can be shared among multiple data sets. The \texttt{ImSim} module (see \ref{sec:imsim}) contains a class \texttt{Multiband} that naturally handles an arbitrary number of data sets, all with their own descriptions (see \ref{sec:data}). The \texttt{Multiband} class shares the same API as the \texttt{ImageModel} class for single images and thus allows to be used with the \texttt{Sampling} and \texttt{Workflow} modules. The non-linear parameters, such as lens model, point source position and light profile shapes are shared among the different bands. The linear parameters, however, are optimized for each band individually. This allows e.g. for different galaxy morphologies in different wavelength. The multi-band approach allows also to model a set of single exposures directly rather than rely on combined post-processed data products. This approach can also be used to model disjoint patches of a cluster arc without requiring a large image.

Precise relative astrometry may be required to perform the lens modelling in a joint coordinate frame. \pythonpackage\ comes with an iterative routine to align coordinate frames from different bands given a shared model description. This can be used to align images to determine e.g. a point source or the lensing galaxy light center.


\section{Science applications of \pythonpackage}
\label{sec:applications}

In this section we provide science examples that \pythonpackage\ has enabled. In particular, we will highlight specific settings within \pythonpackage\ that were required to conduct the analysis in the domain of substructure lensing, time-delay cosmography and cosmic shear measurements.

As the size and diversity of known strong lensing systems increases, a wider variety of science topics can be tackled, such as time-delay cosmography with lensed SNIa \cite{Grillo:2018}, single star micro-lensing cluster arcs \cite{Kelly:2018}, double source-plane cosmography \cite{Gavazzi:2008, Collett:2014} or cosmic shear measurements with Einstein rings \citep{Birrer2018_cosmic_shear}.

We emphasize that the choices when modelling a specific system remains the task of the user. \pythonpackage\ may facilitate scientific analysis of strong lenses, but should be accompanied by rigorous testing of the specific method applied, desirably through simulations. \pythonpackage\ allows to simulate accurate mock data with very complex structure in lens and source and therefore facilitate the exploration of systematics in the analysis.

\subsection{Lensing substructure quantification}

Modeling substructure within a deflector can be done by combining multiple lens models (e.g. a main deflector, external shear and a small clump) within one instance of \texttt{LensModel} (Section \ref{sec:lens_model}). Substructure can be represented by a NFW profile, truncated NFW profile or a variety of other profiles implemented in \pythonpackage. A superposition of an arbitrary number of lens profiles based on a mass function is possible and has been used by \cite{Gilman:20117_abc}. The multi-plane setting also allows to model the full line-of-sight contribution of field halos.

Lensing substructure is expected to perturb the deflection angles at the milliarcsecond scale, which is  e.g. below the pixel resolution of an HST image. To detect and/or quantify those astrometric anomalies, the numerical description in the modelling must accurately capture these small effects. Sub-grid resolution ray-tracing is required to perform such analysis on HST images. Accuracy comes with a computational cost and \pythonpackage\ enables the user to set the right numerical description for the problem in hand.

Additionally, the source surface brightness resolution captured by the model must be sufficiently high resolution not to falsely attribute residuals in the image reconstruction to lensing substructure when they originate from missing scales in the source reconstruction. In \cite{Birrer:2017a}, we specifically enhanced the source reconstruction resolution where we proposed a clump to be present.

\subsection{Time-delay cosmography}
The workflow API facilitates a fast exploration of various choices and options in all the aspects of lens modelling. It is necessary to explore the degeneracies inherent in lensing and their impact on the cosmographic inference. In \cite{Birrer2016_mst}, we explored the source scale degeneracy by explicitly mapping out the source size with the shapelet scale parameter. In \cite{Birrer:J1206} we combined 128 different model settings based on their relative Bayesian Information Criteria to provide a posterior distribution reflecting uncertainties in the model choices. The built-in time-delay likelihood and the \texttt{GalKin} module provide the full support for a fully self-consistent analysis of imaging, time-delay and kinematic data to derive cosmographic constraints.

\subsection{Cosmic shear measurements}
We applied \pythonpackage\ to model and reconstruct the non-linear shear distortions that couple to the main deflector in an Einstein ring lens in the COSMOS field \citep{Birrer_2017los}. The detailed modelling of the HST imaging of the Einstein ring allowed us to constrain the shear parameters to very high precision. \pythonpackage\ is aimed to have the flexibility to model hundreds or even the aimed thousands of Einstein ring lenses expected in future space based surveys to provide comparable and complementary cosmic shear measurements, as been fore-casted by \cite{Birrer2018_cosmic_shear}.

\section{Conclusion} \label{sec:conclusion}
We have presented \pythonpackage\ , a multi-purpose open source lens modelling software package in \Python. We outlined its design and the major supported features. \pythonpackage\ has been used to study the expansion history of the universe with time-delay cosmography and to probe dark matter properties by substructure lensing. The modular nature of \pythonpackage\ provides support for a wide range of scientific studies. We have provided modelling and science examples to illustrate some of the capabilities of \pythonpackage\ . The software is distributed under the MIT license. The software is actively used and maintained and the latest stable release will be distributed through the python packaging index. We refer to the online documentation\footnote{\docLink}, where the latest starter guide, example notebooks, source code and installation guidelines can be found.

\section*{Acknowledgements}
SB thanks Alexandre Refregier and Tommaso Treu for useful comments and support that enabled the development and public release of \pythonpackage. SB thanks Joel Akeret for valuable software engineering support and advices in design and good practice software development. The software was and is actively in use by multiple users. We especially thank Anowar Shajib, Daniel Gilman, Felix A. Kuhn, Kevin Fusshoeller, Cyril Welschen, Felix Mayor, Remy Joseph, Martin Millon, Xuheng Ding and Brian Nord. Their valuable feedback lead to improvements in stability, bug fixing, more general design implementations and extended the capabilities of \pythonpackage.

SB acknowledges support from NASA through grant HST-GO-14254 and HST-GO-14630 from the Space Telescope Science Institute, which is operated by the Association of Universities for Research in Astronomy, Inc., under NASA contract NAS 5-26555.

\appendix
\section{Publicly available lens modelling software} \label{app:software}
A collection of public available lens modelling software presented in the literature is listed below. We refer to specific literature and online documentations for the scope of each individual software and its current development status.

\begin{itemize}
    \item \texttt{gravlens} \citep{gravlens_ascl}\footnote{
\url{http://www.physics.rutgers.edu/~keeton/gravlens/}}: A standard lens model software widely used in the community. Includes a wide range of basic lensing calculations and comes with an extension that adds many routines for modeling strong lenses.
    \item \texttt{lenstool} \citep{lenstool_ascl}\footnote{\url{
http://projets.lam.fr/projects/lenstool/wiki}}: A lensing software for modeling mass distribution of galaxies and clusters. Comes with a Bayesian inference method.
    \item \texttt{PixeLens} \citep{pixeLens_ascl}\footnote{\url{http://www.physik.uzh.ch/~psaha/lens/pixelens.php}}: A program for reconstructing gravitational lenses from multiple-imaged point sources. It can explore ensembles of lens-models consistent with given data on several different lens systems at once.
    \item \texttt{glafic} \citep{Oguri:2010}\footnote{\url{
http://www.slac.stanford.edu/~oguri/glafic/}}: Support for many mass models and parametric light models. Simulats lensed extended images with PSF convolution.
    \item \texttt{LENSED} \citep{Tessore:2016}\footnote{\url{
http://glenco.github.io/lensed/}}: Performs forward parametric modelling of strong lenses. Supports computing on GPUs.
    \item \texttt{AutoLens} \citep{Nightingale:2018}\footnote{\url{http://jamesnightingale.net/AutoLens/}}: An automated modeling suite for the analysis of galaxy-scale strong gravitational lenses. Incorporates an adaptive grid source reconstruction technique.
    \item \texttt{Ensai} \citep{Hezaveh:2017}\footnote{\url{https://github.com/yasharhezaveh/Ensai}}: Estimating parameters of strong gravitational lenses with convolutional neural networks.
    \item \texttt{pySPT} \citep{Wertz2018_pySPT}\footnote{\url{https://github.com/owertz/pySPT}}: A package dedicated to the Source Position Transformation (SPT). The main goal of pySPT is to provide a tool to quantify the systematic errors that are introduced by the SPT in lens modeling.
\end{itemize}



\bibliographystyle{elsarticle-harv} 
\bibliography{BibdeskLib}

\begin{thebibliography}{64}
\expandafter\ifx\csname natexlab\endcsname\relax\def\natexlab#1{#1}\fi
\expandafter\ifx\csname url\endcsname\relax
  \def\url#1{\texttt{#1}}\fi
\expandafter\ifx\csname urlprefix\endcsname\relax\def\urlprefix{URL }\fi

\bibitem[{{Agnello} et~al.(2015){Agnello}, {Treu}, {Ostrovski}, {Schechter},
  {Buckley-Geer}, {Lin}, {Auger}, {Courbin}, {Fassnacht}, {Frieman},
  {Kuropatkin}, {Marshall}, {McMahon}, {Meylan}, {More}, {Suyu}, {Rusu},
  {Finley}, {Abbott}, {Abdalla}, {Allam}, {Annis}, {Banerji},
  {Benoit-L{\'e}vy}, {Bertin}, {Brooks}, {Burke}, {Carnero Rosell}, {Carrasco
  Kind}, {Carretero}, {Cunha}, {D'Andrea}, {da Costa}, {Desai}, {Diehl},
  {Dietrich}, {Doel}, {Eifler}, {Estrada}, {Fausti Neto}, {Flaugher},
  {Fosalba}, {Gerdes}, {Gruen}, {Gutierrez}, {Honscheid}, {James}, {Kuehn},
  {Lahav}, {Lima}, {Maia}, {March}, {Marshall}, {Martini}, {Melchior},
  {Miller}, {Miquel}, {Nichol}, {Ogando}, {Plazas}, {Reil}, {Romer}, {Roodman},
  {Sako}, {Sanchez}, {Santiago}, {Scarpine}, {Schubnell}, {Sevilla-Noarbe},
  {Smith}, {Soares-Santos}, {Sobreira}, {Suchyta}, {Swanson}, {Tarle},
  {Thaler}, {Tucker}, {Walker}, {Wechsler}, and {Zhang}}]{Agnello:2015DES}
{Agnello}, A., {Treu}, T., {Ostrovski}, F., {Schechter}, P.~L., {Buckley-Geer},
  E.~J., {Lin}, H., {Auger}, M.~W., {Courbin}, F., {Fassnacht}, C.~D.,
  {Frieman}, J., {Kuropatkin}, N., {Marshall}, P.~J., {McMahon}, R.~G.,
  {Meylan}, G., {More}, A., {Suyu}, S.~H., {Rusu}, C.~E., {Finley}, D.,
  {Abbott}, T., {Abdalla}, F.~B., {Allam}, S., {Annis}, J., {Banerji}, M.,
  {Benoit-L{\'e}vy}, A., {Bertin}, E., {Brooks}, D., {Burke}, D.~L., {Carnero
  Rosell}, A., {Carrasco Kind}, M., {Carretero}, J., {Cunha}, C.~E.,
  {D'Andrea}, C.~B., {da Costa}, L.~N., {Desai}, S., {Diehl}, H.~T.,
  {Dietrich}, J.~P., {Doel}, P., {Eifler}, T.~F., {Estrada}, J., {Fausti Neto},
  A., {Flaugher}, B., {Fosalba}, P., {Gerdes}, D.~W., {Gruen}, D., {Gutierrez},
  G., {Honscheid}, K., {James}, D.~J., {Kuehn}, K., {Lahav}, O., {Lima}, M.,
  {Maia}, M.~A.~G., {March}, M., {Marshall}, J.~L., {Martini}, P., {Melchior},
  P., {Miller}, C.~J., {Miquel}, R., {Nichol}, R.~C., {Ogando}, R., {Plazas},
  A.~A., {Reil}, K., {Romer}, A.~K., {Roodman}, A., {Sako}, M., {Sanchez}, E.,
  {Santiago}, B., {Scarpine}, V., {Schubnell}, M., {Sevilla-Noarbe}, I.,
  {Smith}, R.~C., {Soares-Santos}, M., {Sobreira}, F., {Suchyta}, E.,
  {Swanson}, M.~E.~C., {Tarle}, G., {Thaler}, J., {Tucker}, D., {Walker},
  A.~R., {Wechsler}, R.~H., {Zhang}, Y., Dec. 2015. {Discovery of two
  gravitationally lensed quasars in the Dark Energy Survey}. \mnras 454,
  1260--1265.

\bibitem[{{Akeret} et~al.(2013){Akeret}, {Seehars}, {Amara}, {Refregier}, and
  {Csillaghy}}]{Akeret:2013nl}
{Akeret}, J., {Seehars}, S., {Amara}, A., {Refregier}, A., {Csillaghy}, A.,
  Aug. 2013. {CosmoHammer: Cosmological parameter estimation with the MCMC
  Hammer}. Astronomy and Computing 2, 27--39.

\bibitem[{{Astropy Collaboration} et~al.(2013){Astropy Collaboration},
  {Robitaille}, {Tollerud}, {Greenfield}, {Droettboom}, {Bray}, {Aldcroft},
  {Davis}, {Ginsburg}, {Price-Whelan}, {Kerzendorf}, {Conley}, {Crighton},
  {Barbary}, {Muna}, {Ferguson}, {Grollier}, {Parikh}, {Nair}, {Unther},
  {Deil}, {Woillez}, {Conseil}, {Kramer}, {Turner}, {Singer}, {Fox}, {Weaver},
  {Zabalza}, {Edwards}, {Azalee Bostroem}, {Burke}, {Casey}, {Crawford},
  {Dencheva}, {Ely}, {Jenness}, {Labrie}, {Lim}, {Pierfederici}, {Pontzen},
  {Ptak}, {Refsdal}, {Servillat}, and {Streicher}}]{astropy}
{Astropy Collaboration}, {Robitaille}, T.~P., {Tollerud}, E.~J., {Greenfield},
  P., {Droettboom}, M., {Bray}, E., {Aldcroft}, T., {Davis}, M., {Ginsburg},
  A., {Price-Whelan}, A.~M., {Kerzendorf}, W.~E., {Conley}, A., {Crighton}, N.,
  {Barbary}, K., {Muna}, D., {Ferguson}, H., {Grollier}, F., {Parikh}, M.~M.,
  {Nair}, P.~H., {Unther}, H.~M., {Deil}, C., {Woillez}, J., {Conseil}, S.,
  {Kramer}, R., {Turner}, J.~E.~H., {Singer}, L., {Fox}, R., {Weaver}, B.~A.,
  {Zabalza}, V., {Edwards}, Z.~I., {Azalee Bostroem}, K., {Burke}, D.~J.,
  {Casey}, A.~R., {Crawford}, S.~M., {Dencheva}, N., {Ely}, J., {Jenness}, T.,
  {Labrie}, K., {Lim}, P.~L., {Pierfederici}, F., {Pontzen}, A., {Ptak}, A.,
  {Refsdal}, B., {Servillat}, M., {Streicher}, O., Oct. 2013. {Astropy: A
  community Python package for astronomy}. \aap 558, A33.

\bibitem[{{Birrer} et~al.(2015){Birrer}, {Amara}, and
  {Refregier}}]{Birrer2015_basis_set}
{Birrer}, S., {Amara}, A., {Refregier}, A., Nov. 2015. {Gravitational Lens
  Modeling with Basis Sets}. \apj 813, 102.

\bibitem[{{Birrer} et~al.(2016){Birrer}, {Amara}, and
  {Refregier}}]{Birrer2016_mst}
{Birrer}, S., {Amara}, A., {Refregier}, A., Aug. 2016. {The mass-sheet
  degeneracy and time-delay cosmography: analysis of the strong lens
  RXJ1131-1231}. Journal of Cosmology and Astro-Particle Physics 2016, 020.

\bibitem[{{Birrer} et~al.(2017{\natexlab{a}}){Birrer}, {Amara}, and
  {Refregier}}]{Birrer:2017a}
{Birrer}, S., {Amara}, A., {Refregier}, A., May 2017{\natexlab{a}}. {Lensing
  substructure quantification in RXJ1131-1231: a 2 keV lower bound on dark
  matter thermal relic mass}. Journal of Cosmology and Astro-Particle Physics
  2017, 037.

\bibitem[{{Birrer} et~al.(2018{\natexlab{a}}){Birrer}, {Refregier}, and
  {Amara}}]{Birrer2018_cosmic_shear}
{Birrer}, S., {Refregier}, A., {Amara}, A., Jan. 2018{\natexlab{a}}. {Cosmic
  Shear with Einstein Rings}. \apjl 852, L14.

\bibitem[{{Birrer} et~al.(2018{\natexlab{b}}){Birrer}, {Treu}, {Rusu},
  {Bonvin}, {Fassnacht}, {Chan}, {Agnello}, {Shajib}, {Chen}, {Auger},
  {Courbin}, {Hilbert}, {Sluse}, {Suyu}, {Wong}, {Marshall}, {Lemaux}, and
  {Meylan}}]{Birrer:J1206}
{Birrer}, S., {Treu}, T., {Rusu}, C.~E., {Bonvin}, V., {Fassnacht}, C.~D.,
  {Chan}, J.~H.~H., {Agnello}, A., {Shajib}, A.~J., {Chen}, G.~C.~F., {Auger},
  M., {Courbin}, F., {Hilbert}, S., {Sluse}, D., {Suyu}, S.~H., {Wong}, K.~C.,
  {Marshall}, P., {Lemaux}, B.~C., {Meylan}, G., Sep. 2018{\natexlab{b}}.
  {H0LiCOW - IX. Cosmographic analysis of the doubly imaged quasar SDSS
  1206+4332 and a new measurement of the Hubble constant}. ArXiv e-prints,
  arXiv:1809.01274.

\bibitem[{{Birrer} et~al.(2017{\natexlab{b}}){Birrer}, {Welschen}, {Amara}, and
  {Refregier}}]{Birrer_2017los}
{Birrer}, S., {Welschen}, C., {Amara}, A., {Refregier}, A., Apr.
  2017{\natexlab{b}}. {Line-of-sight effects in strong lensing: putting theory
  into practice}. Journal of Cosmology and Astro-Particle Physics 2017, 049.

\bibitem[{{Bonvin} et~al.(2017){Bonvin}, {Courbin}, {Suyu}, {Marshall}, {Rusu},
  {Sluse}, {Tewes}, {Wong}, {Collett}, {Fassnacht}, {Treu}, {Auger}, {Hilbert},
  {Koopmans}, {Meylan}, {Rumbaugh}, {Sonnenfeld}, and
  {Spiniello}}]{Bonvin:2017}
{Bonvin}, V., {Courbin}, F., {Suyu}, S.~H., {Marshall}, P.~J., {Rusu}, C.~E.,
  {Sluse}, D., {Tewes}, M., {Wong}, K.~C., {Collett}, T., {Fassnacht}, C.~D.,
  {Treu}, T., {Auger}, M.~W., {Hilbert}, S., {Koopmans}, L.~V.~E., {Meylan},
  G., {Rumbaugh}, N., {Sonnenfeld}, A., {Spiniello}, C., Mar. 2017. {H0LiCOW -
  V. New COSMOGRAIL time delays of HE 0435-1223: H$_{0}$ to 3.8 per cent
  precision from strong lensing in a flat {$\Lambda$}CDM model}. \mnras 465,
  4914--4930.

\bibitem[{{Cappellari}(2002)}]{Cappellari:2002_mge}
{Cappellari}, M., Jun. 2002. {Efficient multi-Gaussian expansion of galaxies}.
  \mnras 333, 400--410.

\bibitem[{{Collett} and {Auger}(2014)}]{Collett:2014}
{Collett}, T.~E., {Auger}, M.~W., Sep. 2014. {Cosmological constraints from the
  double source plane lens SDSSJ0946+1006}. \mnras 443, 969--976.

\bibitem[{{Dalal} and {Kochanek}(2002)}]{Dalal:2002}
{Dalal}, N., {Kochanek}, C.~S., Jun. 2002. {Direct Detection of Cold Dark
  Matter Substructure}. \apj 572, 25--33.

\bibitem[{{Ding} et~al.(2018){Ding}, {Treu}, {Shajib}, {Xu}, {Chen}, {More},
  {Despali}, {Frigo}, {Fassnacht}, {Gilman}, {Hilbert}, {Marshall}, {Sluse},
  and {Vegetti}}]{tdlmc:2018}
{Ding}, X., {Treu}, T., {Shajib}, A.~J., {Xu}, D., {Chen}, G.~C.-F., {More},
  A., {Despali}, G., {Frigo}, M., {Fassnacht}, C.~D., {Gilman}, D., {Hilbert},
  S., {Marshall}, P.~J., {Sluse}, D., {Vegetti}, S., Jan. 2018. {Time Delay
  Lens Modeling Challenge: I. Experimental Design}. ArXiv e-prints.

\bibitem[{{Foreman-Mackey} et~al.(2013){Foreman-Mackey}, {Hogg}, {Lang}, and
  {Goodman}}]{emcee}
{Foreman-Mackey}, D., {Hogg}, D.~W., {Lang}, D., {Goodman}, J., Mar. 2013.
  {emcee: The MCMC Hammer}. \pasp 125, 306.

\bibitem[{{Gavazzi} et~al.(2008){Gavazzi}, {Treu}, {Koopmans}, {Bolton},
  {Moustakas}, {Burles}, and {Marshall}}]{Gavazzi:2008}
{Gavazzi}, R., {Treu}, T., {Koopmans}, L.~V.~E., {Bolton}, A.~S., {Moustakas},
  L.~A., {Burles}, S., {Marshall}, P.~J., Apr. 2008. {The Sloan Lens ACS
  Survey. VI. Discovery and Analysis of a Double Einstein Ring}. \apj 677,
  1046--1059.

\bibitem[{{Gilman} et~al.(2018){Gilman}, {Birrer}, {Treu}, {Keeton}, and
  {Nierenberg}}]{Gilman:20117_abc}
{Gilman}, D., {Birrer}, S., {Treu}, T., {Keeton}, C.~R., {Nierenberg}, A., Nov.
  2018. {Probing the nature of dark matter by forward modelling flux ratios in
  strong gravitational lenses}. \mnras 481, 819--834.

\bibitem[{{Grillo} et~al.(2018){Grillo}, {Rosati}, {Suyu}, {Balestra},
  {Caminha}, {Halkola}, {Kelly}, {Lombardi}, {Mercurio}, {Rodney}, and
  {Treu}}]{Grillo:2018}
{Grillo}, C., {Rosati}, P., {Suyu}, S.~H., {Balestra}, I., {Caminha}, G.~B.,
  {Halkola}, A., {Kelly}, P.~L., {Lombardi}, M., {Mercurio}, A., {Rodney},
  S.~A., {Treu}, T., Jun. 2018. {Measuring the Value of the Hubble Constant
  {\textquotedblleft}{\`a} la Refsdal{\textquotedblright}}. \apj 860, 94.

\bibitem[{{Hezaveh} et~al.(2016){Hezaveh}, {Dalal}, {Marrone}, {Mao},
  {Morningstar}, {Wen}, {Blandford}, {Carlstrom}, {Fassnacht}, {Holder},
  {Kemball}, {Marshall}, {Murray}, {Perreault Levasseur}, {Vieira}, and
  {Wechsler}}]{Hezaveh:2016uj}
{Hezaveh}, Y.~D., {Dalal}, N., {Marrone}, D.~P., {Mao}, Y.-Y., {Morningstar},
  W., {Wen}, D., {Blandford}, R.~D., {Carlstrom}, J.~E., {Fassnacht}, C.~D.,
  {Holder}, G.~P., {Kemball}, A., {Marshall}, P.~J., {Murray}, N., {Perreault
  Levasseur}, L., {Vieira}, J.~D., {Wechsler}, R.~H., May 2016. {Detection of
  Lensing Substructure Using ALMA Observations of the Dusty Galaxy SDP.81}.
  \apj 823, 37.

\bibitem[{{Hezaveh} et~al.(2017){Hezaveh}, {Levasseur}, and
  {Marshall}}]{Hezaveh:2017}
{Hezaveh}, Y.~D., {Levasseur}, L.~P., {Marshall}, P.~J., Aug. 2017. {Fast
  automated analysis of strong gravitational lenses with convolutional neural
  networks}. \nat 548, 555--557.

\bibitem[{Hunter(2007)}]{matplotlib}
Hunter, J.~D., 2007. Matplotlib: A 2d graphics environment. Computing In
  Science \& Engineering 9~(3), 90--95.

\bibitem[{{Jacobs} et~al.(2017){Jacobs}, {Glazebrook}, {Collett}, {More}, and
  {McCarthy}}]{Jacobs:2017}
{Jacobs}, C., {Glazebrook}, K., {Collett}, T., {More}, A., {McCarthy}, C., Oct.
  2017. {Finding strong lenses in CFHTLS using convolutional neural networks}.
  \mnras 471, 167--181.

\bibitem[{{Joseph} et~al.(2018){Joseph}, {Courbin}, {Starck}, and
  {Birrer}}]{Joseph:2018}
{Joseph}, R., {Courbin}, F., {Starck}, J.~L., {Birrer}, S., Sep. 2018. {Sparse
  Lens Inversion Technique (SLIT): lens and source separability from linear
  inversion of the source reconstruction problem}. ArXiv e-prints,
  arXiv:1809.09121.

\bibitem[{{Keeton}(2011)}]{gravlens_ascl}
{Keeton}, C.~R., Feb. 2011. {GRAVLENS: Computational Methods for Gravitational
  Lensing}. Astrophysics Source Code Library.

\bibitem[{{Kelly} et~al.(2018){Kelly}, {Diego}, {Rodney}, {Kaiser},
  {Broadhurst}, {Zitrin}, {Treu}, {P{\'e}rez-Gonz{\'a}lez}, {Morishita},
  {Jauzac}, {Selsing}, {Oguri}, {Pueyo}, {Ross}, {Filippenko}, {Smith},
  {Hjorth}, {Cenko}, {Wang}, {Howell}, {Richard}, {Frye}, {Jha}, {Foley},
  {Norman}, {Bradac}, {Zheng}, {Brammer}, {Benito}, {Cava}, {Christensen}, {de
  Mink}, {Graur}, {Grillo}, {Kawamata}, {Kneib}, {Matheson}, {McCully},
  {Nonino}, {P{\'e}rez-Fournon}, {Riess}, {Rosati}, {Schmidt}, {Sharon}, and
  {Weiner}}]{Kelly:2018}
{Kelly}, P.~L., {Diego}, J.~M., {Rodney}, S., {Kaiser}, N., {Broadhurst}, T.,
  {Zitrin}, A., {Treu}, T., {P{\'e}rez-Gonz{\'a}lez}, P.~G., {Morishita}, T.,
  {Jauzac}, M., {Selsing}, J., {Oguri}, M., {Pueyo}, L., {Ross}, T.~W.,
  {Filippenko}, A.~V., {Smith}, N., {Hjorth}, J., {Cenko}, S.~B., {Wang}, X.,
  {Howell}, D.~A., {Richard}, J., {Frye}, B.~L., {Jha}, S.~W., {Foley}, R.~J.,
  {Norman}, C., {Bradac}, M., {Zheng}, W., {Brammer}, G., {Benito}, A.~M.,
  {Cava}, A., {Christensen}, L., {de Mink}, S.~E., {Graur}, O., {Grillo}, C.,
  {Kawamata}, R., {Kneib}, J.-P., {Matheson}, T., {McCully}, C., {Nonino}, M.,
  {P{\'e}rez-Fournon}, I., {Riess}, A.~G., {Rosati}, P., {Schmidt}, K.~B.,
  {Sharon}, K., {Weiner}, B.~J., Apr. 2018. {Extreme magnification of an
  individual star at redshift 1.5 by a galaxy- cluster lens}. Nature Astronomy
  2, 334--342.

\bibitem[{Kennedy and Eberhart(1995)}]{Eberhart:1995qm}
Kennedy, J., Eberhart, R., 01 1995. A new optimizer using particle swarm
  theory. In: Proceedings of IEEE International Conference on Neural Networks.
  IV. pp. 1942--1948.
\newline\urlprefix\url{doi:10.1109/ICNN.1995.488968}

\bibitem[{{Kneib} et~al.(2011){Kneib}, {Bonnet}, {Golse}, {Sand}, {Jullo}, and
  {Marshall}}]{lenstool_ascl}
{Kneib}, J.-P., {Bonnet}, H., {Golse}, G., {Sand}, D., {Jullo}, E., {Marshall},
  P., Feb. 2011. {LENSTOOL: A Gravitational Lensing Software for Modeling Mass
  Distribution of Galaxies and Clusters (strong and weak regime)}. Astrophysics
  Source Code Library.

\bibitem[{{Lemon} et~al.(2018){Lemon}, {Auger}, {McMahon}, and
  {Ostrovski}}]{Lemon:2018}
{Lemon}, C.~A., {Auger}, M.~W., {McMahon}, R.~G., {Ostrovski}, F., Mar. 2018.
  {Gravitationally Lensed Quasars in Gaia: II. Discovery of 24 Lensed Quasars}.
  ArXiv e-prints.

\bibitem[{{Lin} et~al.(2017){Lin}, {Buckley-Geer}, {Agnello}, {Ostrovski},
  {McMahon}, {Nord}, {Kuropatkin}, {Tucker}, {Treu}, {Chan}, {Suyu}, {Diehl},
  {Collett}, {Gill}, {More}, {Amara}, {Auger}, {Courbin}, {Fassnacht},
  {Frieman}, {Marshall}, {Meylan}, {Rusu}, {Abbott}, {Abdalla}, {Allam},
  {Banerji}, {Bechtol}, {Benoit-L{\'e}vy}, {Bertin}, {Brooks}, {Burke},
  {Carnero Rosell}, {Carrasco Kind}, {Carretero}, {Castander}, {Crocce},
  {D'Andrea}, {da Costa}, {Desai}, {Dietrich}, {Eifler}, {Finley}, {Flaugher},
  {Fosalba}, {Garc{\'{\i}}a-Bellido}, {Gaztanaga}, {Gerdes}, {Goldstein},
  {Gruen}, {Gruendl}, {Gschwend}, {Gutierrez}, {Honscheid}, {James}, {Kuehn},
  {Lahav}, {Li}, {Lima}, {Maia}, {March}, {Marshall}, {Martini}, {Melchior},
  {Menanteau}, {Miquel}, {Ogando}, {Plazas}, {Romer}, {Sanchez}, {Schindler},
  {Schubnell}, {Sevilla-Noarbe}, {Smith}, {Smith}, {Sobreira}, {Suchyta},
  {Swanson}, {Tarle}, {Thomas}, {Walker}, and {DES
  Collaboration}}]{Lin:2017DES}
{Lin}, H., {Buckley-Geer}, E., {Agnello}, A., {Ostrovski}, F., {McMahon},
  R.~G., {Nord}, B., {Kuropatkin}, N., {Tucker}, D.~L., {Treu}, T., {Chan},
  J.~H.~H., {Suyu}, S.~H., {Diehl}, H.~T., {Collett}, T., {Gill}, M.~S.~S.,
  {More}, A., {Amara}, A., {Auger}, M.~W., {Courbin}, F., {Fassnacht}, C.~D.,
  {Frieman}, J., {Marshall}, P.~J., {Meylan}, G., {Rusu}, C.~E., {Abbott},
  T.~M.~C., {Abdalla}, F.~B., {Allam}, S., {Banerji}, M., {Bechtol}, K.,
  {Benoit-L{\'e}vy}, A., {Bertin}, E., {Brooks}, D., {Burke}, D.~L., {Carnero
  Rosell}, A., {Carrasco Kind}, M., {Carretero}, J., {Castander}, F.~J.,
  {Crocce}, M., {D'Andrea}, C.~B., {da Costa}, L.~N., {Desai}, S., {Dietrich},
  J.~P., {Eifler}, T.~F., {Finley}, D.~A., {Flaugher}, B., {Fosalba}, P.,
  {Garc{\'{\i}}a-Bellido}, J., {Gaztanaga}, E., {Gerdes}, D.~W., {Goldstein},
  D.~A., {Gruen}, D., {Gruendl}, R.~A., {Gschwend}, J., {Gutierrez}, G.,
  {Honscheid}, K., {James}, D.~J., {Kuehn}, K., {Lahav}, O., {Li}, T.~S.,
  {Lima}, M., {Maia}, M.~A.~G., {March}, M., {Marshall}, J.~L., {Martini}, P.,
  {Melchior}, P., {Menanteau}, F., {Miquel}, R., {Ogando}, R.~L.~C., {Plazas},
  A.~A., {Romer}, A.~K., {Sanchez}, E., {Schindler}, R., {Schubnell}, M.,
  {Sevilla-Noarbe}, I., {Smith}, M., {Smith}, R.~C., {Sobreira}, F., {Suchyta},
  E., {Swanson}, M.~E.~C., {Tarle}, G., {Thomas}, D., {Walker}, A.~R., {DES
  Collaboration}, Apr. 2017. {Discovery of the Lensed Quasar System DES
  J0408-5354}. \apjl 838, L15.

\bibitem[{{Mamon} and {{\L}okas}(2005)}]{MamonLokas:2005}
{Mamon}, G.~A., {{\L}okas}, E.~L., Nov. 2005. {Dark matter in elliptical
  galaxies - II. Estimating the mass within the virial radius}. \mnras 363,
  705--722.

\bibitem[{{Mao} and {Schneider}(1998)}]{Mao_Schneider:1998}
{Mao}, S., {Schneider}, P., Apr. 1998. {Evidence for substructure in lens
  galaxies?} \mnras 295, 587.

\bibitem[{{Metcalf} and {Madau}(2001)}]{Metcalf_Madau:2001}
{Metcalf}, R.~B., {Madau}, P., Dec. 2001. {Compound Gravitational Lensing as a
  Probe of Dark Matter Substructure within Galaxy Halos}. \apj 563, 9--20.

\bibitem[{{Nierenberg} et~al.(2017){Nierenberg}, {Treu}, {Brammer}, {Peter},
  {Fassnacht}, {Keeton}, {Kochanek}, {Schmidt}, {Sluse}, and
  {Wright}}]{Nierenberg:2017}
{Nierenberg}, A.~M., {Treu}, T., {Brammer}, G., {Peter}, A.~H.~G., {Fassnacht},
  C.~D., {Keeton}, C.~R., {Kochanek}, C.~S., {Schmidt}, K.~B., {Sluse}, D.,
  {Wright}, S.~A., Oct. 2017. {Probing dark matter substructure in the
  gravitational lens HE 0435-1223 with the WFC3 grism}. \mnras 471, 2224--2236.

\bibitem[{{Nierenberg} et~al.(2014){Nierenberg}, {Treu}, {Wright}, {Fassnacht},
  and {Auger}}]{Nierenberg:2014}
{Nierenberg}, A.~M., {Treu}, T., {Wright}, S.~A., {Fassnacht}, C.~D., {Auger},
  M.~W., Aug. 2014. {Detection of substructure with adaptive optics integral
  field spectroscopy of the gravitational lens B1422+231}. \mnras 442,
  2434--2445.

\bibitem[{{Nightingale} and {Dye}(2015)}]{Nightingale:2015}
{Nightingale}, J.~W., {Dye}, S., Sep. 2015. {Adaptive semi-linear inversion of
  strong gravitational lens imaging}. \mnras 452, 2940--2959.

\bibitem[{{Nightingale} et~al.(2018){Nightingale}, {Dye}, and
  {Massey}}]{Nightingale:2018}
{Nightingale}, J.~W., {Dye}, S., {Massey}, R.~J., Aug. 2018. {AutoLens:
  automated modeling of a strong lens's light, mass, and source}. \mnras 478,
  4738--4784.

\bibitem[{{Nord} et~al.(2016){Nord}, {Buckley-Geer}, {Lin}, {Diehl}, {Helsby},
  {Kuropatkin}, {Amara}, {Collett}, {Allam}, {Caminha}, {De Bom}, {Desai},
  {D{\'u}met-Montoya}, {Pereira}, {Finley}, {Flaugher}, {Furlanetto},
  {Gaitsch}, {Gill}, {Merritt}, {More}, {Tucker}, {Saro}, {Rykoff}, {Rozo},
  {Birrer}, {Abdalla}, {Agnello}, {Auger}, {Brunner}, {Carrasco Kind},
  {Castander}, {Cunha}, {da Costa}, {Foley}, {Gerdes}, {Glazebrook},
  {Gschwend}, {Hartley}, {Kessler}, {Lagattuta}, {Lewis}, {Maia}, {Makler},
  {Menanteau}, {Niernberg}, {Scolnic}, {Vieira}, {Gramillano}, {Abbott},
  {Banerji}, {Benoit-L{\'e}vy}, {Brooks}, {Burke}, {Capozzi}, {Carnero Rosell},
  {Carretero}, {D'Andrea}, {Dietrich}, {Doel}, {Evrard}, {Frieman},
  {Gaztanaga}, {Gruen}, {Honscheid}, {James}, {Kuehn}, {Li}, {Lima},
  {Marshall}, {Martini}, {Melchior}, {Miquel}, {Neilsen}, {Nichol}, {Ogando},
  {Plazas}, {Romer}, {Sako}, {Sanchez}, {Scarpine}, {Schubnell},
  {Sevilla-Noarbe}, {Smith}, {Soares-Santos}, {Sobreira}, {Suchyta}, {Swanson},
  {Tarle}, {Thaler}, {Walker}, {Wester}, {Zhang}, and {DES
  Collaboration}}]{Nord:2016}
{Nord}, B., {Buckley-Geer}, E., {Lin}, H., {Diehl}, H.~T., {Helsby}, J.,
  {Kuropatkin}, N., {Amara}, A., {Collett}, T., {Allam}, S., {Caminha}, G.~B.,
  {De Bom}, C., {Desai}, S., {D{\'u}met-Montoya}, H., {Pereira}, M.~E.~d.~S.,
  {Finley}, D.~A., {Flaugher}, B., {Furlanetto}, C., {Gaitsch}, H., {Gill}, M.,
  {Merritt}, K.~W., {More}, A., {Tucker}, D., {Saro}, A., {Rykoff}, E.~S.,
  {Rozo}, E., {Birrer}, S., {Abdalla}, F.~B., {Agnello}, A., {Auger}, M.,
  {Brunner}, R.~J., {Carrasco Kind}, M., {Castander}, F.~J., {Cunha}, C.~E.,
  {da Costa}, L.~N., {Foley}, R.~J., {Gerdes}, D.~W., {Glazebrook}, K.,
  {Gschwend}, J., {Hartley}, W., {Kessler}, R., {Lagattuta}, D., {Lewis}, G.,
  {Maia}, M.~A.~G., {Makler}, M., {Menanteau}, F., {Niernberg}, A., {Scolnic},
  D., {Vieira}, J.~D., {Gramillano}, R., {Abbott}, T.~M.~C., {Banerji}, M.,
  {Benoit-L{\'e}vy}, A., {Brooks}, D., {Burke}, D.~L., {Capozzi}, D., {Carnero
  Rosell}, A., {Carretero}, J., {D'Andrea}, C.~B., {Dietrich}, J.~P., {Doel},
  P., {Evrard}, A.~E., {Frieman}, J., {Gaztanaga}, E., {Gruen}, D.,
  {Honscheid}, K., {James}, D.~J., {Kuehn}, K., {Li}, T.~S., {Lima}, M.,
  {Marshall}, J.~L., {Martini}, P., {Melchior}, P., {Miquel}, R., {Neilsen},
  E., {Nichol}, R.~C., {Ogando}, R., {Plazas}, A.~A., {Romer}, A.~K., {Sako},
  M., {Sanchez}, E., {Scarpine}, V., {Schubnell}, M., {Sevilla-Noarbe}, I.,
  {Smith}, R.~C., {Soares-Santos}, M., {Sobreira}, F., {Suchyta}, E.,
  {Swanson}, M.~E.~C., {Tarle}, G., {Thaler}, J., {Walker}, A.~R., {Wester},
  W., {Zhang}, Y., {DES Collaboration}, Aug. 2016. {Observation and
  Confirmation of Six Strong-lensing Systems in the Dark Energy Survey Science
  Verification Data}. \apj 827, 51.

\bibitem[{{Oguri}(2010)}]{Oguri:2010}
{Oguri}, M., Aug. 2010. {The Mass Distribution of SDSS J1004+4112 Revisited}.
  \pasj 62, 1017--1024.

\bibitem[{{Ostrovski} et~al.(2017){Ostrovski}, {McMahon}, {Connolly}, {Lemon},
  {Auger}, {Banerji}, {Hung}, {Koposov}, {Lidman}, {Reed}, {Allam},
  {Benoit-L{\'e}vy}, {Bertin}, {Brooks}, {Buckley-Geer}, {Carnero Rosell},
  {Carrasco Kind}, {Carretero}, {Cunha}, {da Costa}, {Desai}, {Diehl},
  {Dietrich}, {Evrard}, {Finley}, {Flaugher}, {Fosalba}, {Frieman}, {Gerdes},
  {Goldstein}, {Gruen}, {Gruendl}, {Gutierrez}, {Honscheid}, {James}, {Kuehn},
  {Kuropatkin}, {Lima}, {Lin}, {Maia}, {Marshall}, {Martini}, {Melchior},
  {Miquel}, {Ogando}, {Plazas Malag{\'o}n}, {Reil}, {Romer}, {Sanchez},
  {Santiago}, {Scarpine}, {Sevilla-Noarbe}, {Soares-Santos}, {Sobreira},
  {Suchyta}, {Tarle}, {Thomas}, {Tucker}, and {Walker}}]{Ostrovski:2017}
{Ostrovski}, F., {McMahon}, R.~G., {Connolly}, A.~J., {Lemon}, C.~A., {Auger},
  M.~W., {Banerji}, M., {Hung}, J.~M., {Koposov}, S.~E., {Lidman}, C.~E.,
  {Reed}, S.~L., {Allam}, S., {Benoit-L{\'e}vy}, A., {Bertin}, E., {Brooks},
  D., {Buckley-Geer}, E., {Carnero Rosell}, A., {Carrasco Kind}, M.,
  {Carretero}, J., {Cunha}, C.~E., {da Costa}, L.~N., {Desai}, S., {Diehl},
  H.~T., {Dietrich}, J.~P., {Evrard}, A.~E., {Finley}, D.~A., {Flaugher}, B.,
  {Fosalba}, P., {Frieman}, J., {Gerdes}, D.~W., {Goldstein}, D.~A., {Gruen},
  D., {Gruendl}, R.~A., {Gutierrez}, G., {Honscheid}, K., {James}, D.~J.,
  {Kuehn}, K., {Kuropatkin}, N., {Lima}, M., {Lin}, H., {Maia}, M.~A.~G.,
  {Marshall}, J.~L., {Martini}, P., {Melchior}, P., {Miquel}, R., {Ogando}, R.,
  {Plazas Malag{\'o}n}, A., {Reil}, K., {Romer}, K., {Sanchez}, E., {Santiago},
  B., {Scarpine}, V., {Sevilla-Noarbe}, I., {Soares-Santos}, M., {Sobreira},
  F., {Suchyta}, E., {Tarle}, G., {Thomas}, D., {Tucker}, D.~L., {Walker},
  A.~R., Mar. 2017. {VDES J2325-5229 a z = 2.7 gravitationally lensed quasar
  discovered using morphology-independent supervised machine learning}. \mnras
  465, 4325--4334.

\bibitem[{{Peng} et~al.(2002){Peng}, {Ho}, {Impey}, and
  {Rix}}]{Peng:2002_galfit}
{Peng}, C.~Y., {Ho}, L.~C., {Impey}, C.~D., {Rix}, H.-W., Jul. 2002. {Detailed
  Structural Decomposition of Galaxy Images}. \aj 124, 266--293.

\bibitem[{{Peng} et~al.(2010){Peng}, {Ho}, {Impey}, and
  {Rix}}]{Peng:2010_galfit}
{Peng}, C.~Y., {Ho}, L.~C., {Impey}, C.~D., {Rix}, H.-W., Jun. 2010. {Detailed
  Decomposition of Galaxy Images. II. Beyond Axisymmetric Models}. \aj 139,
  2097--2129.

\bibitem[{{Refregier}(2003)}]{Refregier:2003eg}
{Refregier}, A., Jan. 2003. {Shapelets - I. A method for image analysis}.
  \mnras 338, 35--47.

\bibitem[{{Refsdal}(1964)}]{Refsdal:1964pi}
{Refsdal}, S., Jan. 1964. {On the possibility of determining Hubble's parameter
  and the masses of galaxies from the gravitational lens effect}. \mnras 128,
  307.

\bibitem[{Rossum(1995)}]{python_1995}
Rossum, G., 1995. Python reference manual. Tech. rep., Amsterdam, The
  Netherlands, The Netherlands.

\bibitem[{{Saha} and {Williams}(2011)}]{pixeLens_ascl}
{Saha}, P., {Williams}, L.~L.~R., Feb. 2011. {PixeLens: A Portable Modeler of
  Lensed Quasars}. Astrophysics Source Code Library.

\bibitem[{{Schechter} et~al.(1997){Schechter}, {Bailyn}, {Barr}, {Barvainis},
  {Becker}, {Bernstein}, {Blakeslee}, {Bus}, {Dressler}, {Falco}, {Fesen},
  {Fischer}, {Gebhardt}, {Harmer}, {Hewitt}, {Hjorth}, {Hurt}, {Jaunsen},
  {Mateo}, {Mehlert}, {Richstone}, {Sparke}, {Thorstensen}, {Tonry}, {Wegner},
  {Willmarth}, and {Worthey}}]{Schechter:1997}
{Schechter}, P.~L., {Bailyn}, C.~D., {Barr}, R., {Barvainis}, R., {Becker},
  C.~M., {Bernstein}, G.~M., {Blakeslee}, J.~P., {Bus}, S.~J., {Dressler}, A.,
  {Falco}, E.~E., {Fesen}, R.~A., {Fischer}, P., {Gebhardt}, K., {Harmer}, D.,
  {Hewitt}, J.~N., {Hjorth}, J., {Hurt}, T., {Jaunsen}, A.~O., {Mateo}, M.,
  {Mehlert}, D., {Richstone}, D.~O., {Sparke}, L.~S., {Thorstensen}, J.~R.,
  {Tonry}, J.~L., {Wegner}, G., {Willmarth}, D.~W., {Worthey}, G., Feb. 1997.
  {The Quadruple Gravitational Lens PG 1115+080: Time Delays and Models}. \apj
  475, L85--L88.

\bibitem[{{Schechter} et~al.(2017){Schechter}, {Morgan}, {Chehade}, {Metcalfe},
  {Shanks}, and {McDonald}}]{Schechter:2017}
{Schechter}, P.~L., {Morgan}, N.~D., {Chehade}, B., {Metcalfe}, N., {Shanks},
  T., {McDonald}, M., May 2017. {First Lensed Quasar Systems from the VST-ATLAS
  Survey: One Quad, Two Doubles, and Two Pairs of Lensless Twins}. \aj 153,
  219.

\bibitem[{{Shajib} et~al.(2018){Shajib}, {Treu}, and {Agnello}}]{Shajib:2018}
{Shajib}, A.~J., {Treu}, T., {Agnello}, A., Jan. 2018. {Improving time-delay
  cosmography with spatially resolved kinematics}. \mnras 473, 210--226.

\bibitem[{{Suyu} et~al.(2018){Suyu}, {Chang}, {Courbin}, and
  {Okumura}}]{Suyu:2018}
{Suyu}, S.~H., {Chang}, T.-C., {Courbin}, F., {Okumura}, T., Aug. 2018.
  {Cosmological Distance Indicators}. \ssr 214, 91.

\bibitem[{{Suyu} et~al.(2010){Suyu}, {Marshall}, {Auger}, {Hilbert},
  {Blandford}, {Koopmans}, {Fassnacht}, and {Treu}}]{Suyu:2010rc}
{Suyu}, S.~H., {Marshall}, P.~J., {Auger}, M.~W., {Hilbert}, S., {Blandford},
  R.~D., {Koopmans}, L.~V.~E., {Fassnacht}, C.~D., {Treu}, T., Mar. 2010.
  {Dissecting the Gravitational lens B1608+656. II. Precision Measurements of
  the Hubble Constant, Spatial Curvature, and the Dark Energy Equation of
  State}. \apj 711, 201--221.

\bibitem[{{Suyu} et~al.(2006){Suyu}, {Marshall}, {Hobson}, and
  {Blandford}}]{Suyu:2006}
{Suyu}, S.~H., {Marshall}, P.~J., {Hobson}, M.~P., {Blandford}, R.~D., Sep.
  2006. {A Bayesian analysis of regularized source inversions in gravitational
  lensing}. \mnras 371, 983--998.

\bibitem[{{Suyu} et~al.(2014){Suyu}, {Treu}, {Hilbert}, {Sonnenfeld}, {Auger},
  {Blandford}, {Collett}, {Courbin}, {Fassnacht}, {Koopmans}, {Marshall},
  {Meylan}, {Spiniello}, and {Tewes}}]{Suyu:2014aq}
{Suyu}, S.~H., {Treu}, T., {Hilbert}, S., {Sonnenfeld}, A., {Auger}, M.~W.,
  {Blandford}, R.~D., {Collett}, T., {Courbin}, F., {Fassnacht}, C.~D.,
  {Koopmans}, L.~V.~E., {Marshall}, P.~J., {Meylan}, G., {Spiniello}, C.,
  {Tewes}, M., Jun. 2014. {Cosmology from Gravitational Lens Time Delays and
  Planck Data}. \apj 788, L35.

\bibitem[{{Tagore} and {Keeton}(2014)}]{Tagore:2014}
{Tagore}, A.~S., {Keeton}, C.~R., Nov. 2014. {Statistical and systematic
  uncertainties in pixel-based source reconstruction algorithms for
  gravitational lensing}. \mnras 445, 694--710.

\bibitem[{{Tessore} et~al.(2016){Tessore}, {Bellagamba}, and
  {Metcalf}}]{Tessore:2016}
{Tessore}, N., {Bellagamba}, F., {Metcalf}, R.~B., Dec. 2016. {LENSED: a code
  for the forward reconstruction of lenses and sources from strong lensing
  observations}. \mnras 463, 3115--3128.

\bibitem[{{Treu} and {Koopmans}(2002)}]{Treu_Koopmans:2002}
{Treu}, T., {Koopmans}, L.~V.~E., Dec. 2002. {The internal structure of the
  lens PG1115+080: breaking degeneracies in the value of the Hubble constant}.
  \mnras 337, L6--L10.

\bibitem[{{Treu} and {Marshall}(2016)}]{Treu:2016}
{Treu}, T., {Marshall}, P.~J., Jul. 2016. {Time delay cosmography}. \aapr 24,
  11.

\bibitem[{{Vegetti} et~al.(2018){Vegetti}, {Despali}, {Lovell}, and
  {Enzi}}]{Vegetti:2018}
{Vegetti}, S., {Despali}, G., {Lovell}, M.~R., {Enzi}, W., Dec. 2018.
  {Constraining sterile neutrino cosmologies with strong gravitational lensing
  observations at redshift z ~ 0.2}. \mnras 481, 3661--3669.

\bibitem[{{Vegetti} and {Koopmans}(2009)}]{Vegetti:2009}
{Vegetti}, S., {Koopmans}, L.~V.~E., Jan. 2009. {Bayesian strong
  gravitational-lens modelling on adaptive grids: objective detection of mass
  substructure in Galaxies}. \mnras 392, 945--963.

\bibitem[{{Vegetti} et~al.(2010){Vegetti}, {Koopmans}, {Bolton}, {Treu}, and
  {Gavazzi}}]{Vegetti:2010mb}
{Vegetti}, S., {Koopmans}, L.~V.~E., {Bolton}, A., {Treu}, T., {Gavazzi}, R.,
  Nov. 2010. {Detection of a dark substructure through gravitational imaging}.
  \mnras 408, 1969--1981.

\bibitem[{{Vegetti} et~al.(2012){Vegetti}, {Lagattuta}, {McKean}, {Auger},
  {Fassnacht}, and {Koopmans}}]{Vegetti:2012au}
{Vegetti}, S., {Lagattuta}, D.~J., {McKean}, J.~P., {Auger}, M.~W.,
  {Fassnacht}, C.~D., {Koopmans}, L.~V.~E., Jan. 2012. {Gravitational detection
  of a low-mass dark satellite galaxy at cosmological distance}. \nat 481,
  341--343.

\bibitem[{{Warren} and {Dye}(2003)}]{Warren:2003eg}
{Warren}, S.~J., {Dye}, S., Jun. 2003. {Semilinear Gravitational Lens
  Inversion}. \apj 590, 673--682.

\bibitem[{{Wertz} and {Orthen}(2018)}]{Wertz2018_pySPT}
{Wertz}, O., {Orthen}, B., Jan. 2018. {pySPT: a package dedicated to the source
  position transformation}. ArXiv e-prints.

\bibitem[{{Williams} et~al.(2018){Williams}, {Agnello}, {Treu}, {Abramson},
  {Anguita}, {Apostolovski}, {Chen}, {Fassnacht}, {Hsueh}, {Lemaux}, {Motta},
  {Oldham}, {Rojas}, {Rusu}, {Shajib}, and {Wang}}]{Williams:2017}
{Williams}, P.~R., {Agnello}, A., {Treu}, T., {Abramson}, L.~E., {Anguita}, T.,
  {Apostolovski}, Y., {Chen}, G.~C.~F., {Fassnacht}, C.~D., {Hsueh}, J.~W.,
  {Lemaux}, B.~C., {Motta}, V., {Oldham}, L., {Rojas}, K., {Rusu}, C.~E.,
  {Shajib}, A.~J., {Wang}, X., Jun. 2018. {Discovery of three strongly lensed
  quasars in the Sloan Digital Sky Survey}. \mnras 477, L70--L74.

\bibitem[{{Xu} et~al.(2015){Xu}, {Sluse}, {Gao}, {Wang}, {Frenk}, {Mao},
  {Schneider}, and {Springel}}]{Xu:2015}
{Xu}, D., {Sluse}, D., {Gao}, L., {Wang}, J., {Frenk}, C., {Mao}, S.,
  {Schneider}, P., {Springel}, V., Mar. 2015. {How well can cold dark matter
  substructures account for the observed radio flux-ratio anomalies}. \mnras
  447, 3189--3206.

\end{thebibliography}





\end{document}